\documentclass[letterpaper,showpacs,onecolumn,eqsecnum,superscriptaddress,nofootinbib]{revtex4}
\usepackage[utf8]{inputenc}
\usepackage{amssymb,amsmath,amsthm,amsfonts,epsfig,times,natbib}
\usepackage{graphicx} 

\usepackage[usenames,dvipsnames]{color}
\usepackage{epstopdf}
\usepackage{amsmath,amssymb}
\usepackage{tensor}
\usepackage{mathtools}
\usepackage{amsbsy}
\usepackage{bm,url}
\usepackage[linktocpage,hidelinks]{hyperref}
\usepackage{soul}
\usepackage{commath}
\usepackage[scaled]{beramono}
\usepackage[T1]{fontenc}
\usepackage{graphicx}
\usepackage[caption=false]{subfig} 
\usepackage[normalem]{ulem}

\usepackage[dvipsnames]{xcolor}
\hypersetup{
    colorlinks,
    linkcolor={red!50!black},
    citecolor={blue!50!black},
    urlcolor={blue!80!black}
}
\usepackage{soul}

\begin{document}
\title{
Shadow of a collapsing star in a regular spacetime}

\author{Daniel Nunez} 
\email{danielnunez@ciencias.unam.mx }
\affiliation{Instituto de Ciencias F\'isicas, Universidad Nacional Aut\'onoma 
de M\'exico, Apartado Postal 48-3, 62251, Cuernavaca, Morelos, M\'exico}
\author{Juan Carlos Degollado} 
\email{jcdegollado@icf.unam.mx}
\affiliation{Instituto de Ciencias F\'isicas, Universidad Nacional Aut\'onoma 
de M\'exico, Apartado Postal 48-3, 62251, Cuernavaca, Morelos, M\'exico}

\begin{abstract}
We describe the dynamical formation of the shadow of a collapsing star in a Hayward spacetime in terms of an observer far away from the center and a free falling observer.
By solving the time-like and light-like radial geodesics 
we determine the angular size of the shadow as a function of time.
We found that the formation of the shadow is a finite process for both observers and its size is affected by the Hayward spacetime parameters. 
We consider several scenarios, from the Schwarzschild limit to an extreme Hayward black hole.

\end{abstract}




\maketitle


\section{Introduction}
\label{sec:introduction}
A black hole traps all incident light without emitting any, leading to the expectation that an observer would perceive a dark region in the sky where the black hole is located. Nevertheless, owing to the significant bending of light caused by the black hole's gravitational field, the dimensions and appearance of this region posses a particular imprint of the black hole whose properties may be inferred from the observation of the shadow.

The Event Horizon Telescope (EHT) Collaboration 
successfully generated an image depicting the shadow of the black hole 
located at the core of the M87 galaxy  in 2019 \cite{EventHorizonTelescope:2019ths,EventHorizonTelescope:2019pgp}. Following significant improvements in technology,
the image of the black hole
situated at the center of our own galaxy \cite{EventHorizonTelescope:2022wkp,EventHorizonTelescope:2022apq,EventHorizonTelescope:2022urf} was presented (See \cite{Vagnozzi:2022moj} for more details). Inspired by these observations a constant interest in exploring various aspects of black hole shadows were renewed.

The formation of the shadow over time and how its apparent angular size from a collapsing star was first investigated by Schneider and Perlick \cite{Schneider:2018hge}.  They modeled a non-transparent collapsing ball of dust in a spherically symmetric spacetime.  
In this scenario the time-dependent behavior of the star's surface is the one described by the Oppenheimer-Snyder spacetime while the space-time outside the ball is Schwarzschild metric \cite{Oppenheimer:1939ue}. Given these considerations, Schneider and Perlick established that the shadow forms within a finite time, implying that an observer situated at any fixed position beyond the Schwarzschild radius would eventually perceive a circular shadow with an angular size described by Synge \cite{Synge:1966okc}.

For a collapsing ball of dust with the outside described by the Schwarzschild metric, the angular radius of the shadow is primarly determined by its mass. Nonetheless,
for another metrics the shadow size also may depend on the 
parameters that characterize the spacetime.
 It is therefore relevant to compute the angular size for other spacetimes and see how different from the ordinary ones are. In this work, we take a step in this direction and
calculate the dependence
on time of the angular size of the shadow 
of a collapsing star in a spacetime with no singularities. 
Following the approach of  Schneider and Perlick to compute the angular size of a collapsing star, in this work 
we consider a model of collapse 
for which the exterior region is the Hayward spacetime.

The Hayward metric was introduced as a model that aims to address the problem of singularities in black hole formation \cite{Hayward:2005gi}. 
In contrast to the singularity that is predicted at the center of a black hole in many standard solutions of general relativity, the Hayward spacetime 
has been used to describe the gravitational collapse of matter, which forms a black hole-like structure with a regular core. This means that the spacetime remains well-behaved at the center, avoiding the singularities associated with classical black holes \cite{Frolov:2016pav}.
Several works have been focused on the properties of the Hayward black hole. For instance, the quasinormal modes \cite{Flachi:2012nv,Lin:2013ofa,Lopez:2018aec}
gravitational lensing \cite{Chiba:2017nml,  Wei:2015qca,Zhao:2017cwk,Kumar:2019pjp,AbhishekChowdhuri:2023ekr}, geodesic motion \cite{Abbas:2014oua,Bautista-Olvera:2019blb} or accretion \cite{Debnath:2015hea}. See also \cite{Carballo-Rubio:2018pmi,Carballo-Rubio:2018jzw} and references herein for other astrophysical scenarios.

To model the collapse, 
we assume that the star is formed by a  dust-like fluid and 
collapses in free fall so that each point on the surface of the star follows a radial time geodesic.
In order to give a description of the formation of the shadow 
we consider two of observers one located far away from the collapse and other who falls freely towards the center of the star.
The organization of the paper is as follows: In section \ref{sec:hayward-st} we describe some properties of the regular Hayward spacetime and provide the equations for null and timelike geodesics. We also introduce the tetrad attached to a static observer and other tetrad to the radial infalling observer. In section \ref{sec:shadow_colapse} we present the formation of the shadow in the course of the collapse and in section \ref{sec:final} we give some final remarks.

\section{Hayward spacetime}
\label{sec:hayward-st}
The Hayward spacetime in Painlevé-Gullstrand
coordinates, which are adopted to the freely falling radial observer
starting from rest at infinity ($T$, $r$, $\theta$, $\phi$), is given by \cite{Perez-Roman:2018hfy}
\begin{equation} \label{eq:metric}
    ds^2 = -f(r)dT^2+2dTdr\sqrt{\frac{2M(r)}{r}}+ dr^2+r^2(d\theta^2+\sin^2\theta d\phi^2) \ , 
\end{equation}
where
\begin{equation}\label{eq:mass}
    f(r) = 1 -\frac{2M(r)}{r} \ , \qquad M(r) = \frac{M_0r^3}{r^3+q^3} \ ,
\end{equation}
$M_0$ is the ADM (Arnowitt-Desser-Misner) mass and $q$ is a length-scale parameter that is of the order of the Planck lenght. This parameter is used to measure the deviation of the Hayward spacetime from the Schwarzschild spacetime.
The metric satisfies
regularity at the origin such that $f(r) \rightarrow 1 + O(r^2)$ and a
Schwarzschild asymptotic behavior at radial infinity.
The Painlevé-Gullstrand coordinates have the noticeable feature that the hypersurfaces of
$T =$ const. are flat 3-manifolds, i.e. the metric induced on them is the  3-Euclidean metric.

The location of a horizon is determined by the condition $f(r) = 0$. The outer $r_{+}$ and inner horizon $r_{-}$ exist for values $0\leq q\leq q_m$ and are given by 
\begin{eqnarray}\label{eq:horizon}
 r_{+}/M_0 = \frac{2}{3}+\frac{4}{3}\cos\left[ \frac{1}{3}\cos^{-1}\left( 1- \frac{27q^2}{16M_0^2} \right) \right] \ , \qquad
   r_{-}/M_0 = \frac{2}{3}-\frac{4}{3}\cos\left[ \frac{1}{3}\cos^{-1}\left( 1- \frac{27q^2}{16M_0^2} \right) +\frac{\pi}{3}\right]
  \ .
\end{eqnarray}
The limiting value ${q_m}$ is
\begin{equation}\label{eq:qmax}
    \frac{q_m}{M_0} =\frac{32^{1/3}}{3} \equiv 1.05827... \ .
\end{equation}
The structure of the Hayward spacetime, apart from the regular behavior at the center, is quite similar to the spacetime of Reissner-Nordstrom: one may have 
a black hole with two horizons if $q<q_m$, a extreme black hole if $q=q_m$ and a spacetime with no horizons if $q>q_m$.
In the rest of the work we focus on spacetimes that represent a black hole.

\subsection{Geodesics}

In order to describe a collapsing star in a regular spacetime, we assume that each point on the surface follows a radial geodesic on the Hayward spacetime \eqref{eq:metric} and to prevent the escape of photons we assume a dark star.
The description of the shadow will be given in terms of two different observers; a stationary observer both inside and outside the star's radius $r_s$ and a free falling observer who is always outside the star.

The lagrangian for a massive particle moving in the spacetime \eqref{eq:metric} is given by
\begin{equation}
    \mathcal{L}_m=\frac{1}{2}\left[-\left(1-\frac{2M(r)}{r}\right)\left(\frac{dT}{d\tau}\right)^2+2\sqrt{\frac{2M(r)}{r}}\frac{dT}{d\tau}\frac{dr}{d\tau}+\left(\frac{dr}{d\tau}\right)^2+r^2\left(\frac{d\theta}{d\tau}\right)^2+r^2\sin^2{\theta}\left(\frac{d\varphi}{d\tau}\right)^2\right] \ .
    \label{laggul2}
\end{equation}
The conserved quantities associated to $t$ and $\phi$ are respectively:
\begin{equation}
    \varepsilon=\left(1-\frac{2M(r)}{r}\right)\frac{dT}{d\tau}-\sqrt{\frac{2M(r)}{r}}\frac{dr}{d\tau} \ ,
    \label{epsilonpg}
\end{equation}
\begin{equation}
    \ell=r^2\sin^2{\theta}\frac{d\varphi}{d\tau}
    \label{ellpg} \ ,
\end{equation}
where $\varepsilon$ is the energy of the particle and $\ell$ is the azimuthal angular momentum.

Given the normalization condition of the four velocity for massive particles $u^{\mu}u_\mu=-1$, one gets that in the equatorial plane
\begin{equation}
    1=\left(1-\frac{2M(r)}{r}\right)\left(\frac{dT}{d\tau}\right)^2-2\sqrt{\frac{2M(r)}{r}}\frac{dT}{d\tau}\frac{dr}{d\tau}-\left(\frac{dr}{d\tau}\right)^2 \ ,
    \label{Trtau}
\end{equation}
In the following, we consider the geodesics 
oriented towards the future with respect to the coordinate time $t$ so that $dT/d\tau>0$.
By solving Eq.~\eqref{epsilonpg} and Eq.~\eqref{Trtau} in terms of $\frac{dr}{d\tau}$ and $\frac{dT}{d\tau}$  
one gets
\begin{equation}
    \frac{dr}{d\tau}=-\sqrt{\varepsilon^2-1+\frac{2M(r)}{r}}\ ,
    \label{rtaus}
\end{equation}
and
\begin{equation}
    \frac{dT}{d\tau}=\frac{\varepsilon-\sqrt{2M(r)/r}\sqrt{\varepsilon^2-1+2M(r)/r}}{1-2M(r)/r} \ .
    \label{ttaus}
\end{equation}
On the other hand, the lagrangian for massless particles moving in the Hayward spacetime is
\begin{equation}
    \mathcal{L}_n=\frac{1}{2}\left[-\left(1-\frac{2M(r)}{r}\right)\left(\frac{dT}{d\lambda}\right)^2+2\sqrt{\frac{2M(r)}{r}}\frac{dT}{d\lambda}\frac{dr}{d\lambda}+\left(\frac{dr}{d\lambda}\right)^2+r^2\left(\frac{d\theta}{d\lambda}\right)^2+r^2\sin^2{\theta}\left(\frac{d\varphi}{d\lambda}\right)^2\right]
    \label{eq:lagnull}
\end{equation}
where $\lambda$ denotes an affine parameter.
Similarly to massive particles,  there are two constants of motion associated to the energy $E$, and angular momentum $L_z$.
\begin{equation}    \label{eq:e_null}
    E=\left(1-\frac{2M(r)}{r}\right)\frac{dT}{d\lambda}-\sqrt{\frac{2M(r)}{r}}\frac{dr}{d\lambda} \ ,
\end{equation}
and
\begin{equation}    \label{eq:l_null}
    L_z=r^2\frac{d\varphi}{d\lambda} \sin^2\theta \ .
\end{equation}
In the equatorial plane, the trajectories of massless particles satisfy
\begin{equation}
    0=-\left(1-\frac{2M(r)}{r}\right)\left(\frac{dT}{d\lambda}\right)^2+2\sqrt{\frac{2M(r)}{r}}\frac{dT}{d\lambda}\frac{dr}{d\lambda}+\left(\frac{dr}{d\lambda}\right)^2+r^2\left(\frac{d\varphi}{d\lambda}\right)^2
    \label{eq:laggeotiem}
\end{equation}
dividing by $\left(\frac{d\varphi}{d\lambda}\right)^2$ in equation \eqref{eq:laggeotiem} and using the conserved quantities \eqref{eq:e_null} and \eqref{eq:l_null} one obtains
\begin{equation}
    \left(\frac{dr}{d\varphi}\right)^2=\frac{E^2r^4}{L_z^2}-r^2+2M(r)r \ .
    \label{eq:drdphi}
\end{equation}
Taking the derivative of Eq. \eqref{eq:drdphi} with respect to $\varphi$ 
\begin{equation}
    \frac{d^2r}{d\varphi^2}=2\frac{E^2r^3}{L_z^2} -r + M(r) + rM'(r)\ .
\end{equation}
Given for $M(r)$ in Eq. \eqref{eq:mass} with derivative
\begin{eqnarray}
   M'(r)= \frac{3 M0 q^3 r^2}{(q^3 + r^3)^2} \ ,
\end{eqnarray}
one gets, after some algebra
\begin{equation}
    \frac{d^2r}{d\varphi^2}=\frac{E^2r^3}{L_z^2} -r + 4M(r) - 3\frac{M^2(r)}{M_0}\ .
\end{equation}
The condition $\frac{dr}{d\varphi}=0$ at $r=r_m$ fixes the value of $\frac{E^2}{L_z^2}$ at that radius
\begin{equation}
    \frac{E^2r_m^3}{L_z^2}-r_m+2M(r_m)=0 \ .
    \label{eqrm}
\end{equation}
With the condition $\frac{d^2r}{d\varphi^2}\left|_{r=r_{mp}}=0\right.$ a circular orbit satisfies the condition
\begin{equation}
    r_{mp}-\frac{3M^2(r_{mp})}{M_0}= 0  \ ,
    \label{rm}
\end{equation}
The radius $r_{mp}$ characterizes a sphere whose surface is composed of photons in unstable circular motion.
By using~\eqref{eq:drdphi} in equation~\eqref{eq:laggeotiem}, we arrive at:
\begin{equation}
    \frac{dT}{dr}=\frac{\sqrt{2M(r)r}}{r-2M(r)}\pm\frac{E\sqrt{r^5}}{(r-2M(r))\sqrt{E^2r^3-(r-2M(r))L_z^2}}
    \label{tr}
\end{equation}
In the domain $r>2M(r)$, the second term on the right hand side is larger than the first, thus, in order to have $dT/dr>0$ the plus sign has to be considered.
For massless particles with an extremum at $r=r_m$ given by equation~\eqref{eqrm}, equation~\eqref{eq:drdphi} and~\eqref{tr} become:
\begin{equation}
    \left(\frac{dr}{d\varphi}\right)^2=\frac{r^4(r_m-2M(r_m))}{r_m^3}-r^2+2M(r)r \ ,
    \label{rphirm}
\end{equation}
and
\begin{equation}
    \frac{dT}{dr}=\frac{\sqrt{2M(r)r}}{r-2M(r)}\pm\frac{\sqrt{r^5(r_m-2M(r_m))}}{(r-2M(r))\sqrt{(r_m-2M(r_m))r^3-(r-2M(r))r_m^3}} \ .
    \label{ttray}
\end{equation}

The purpose of the following section is to determine the angular radius of this shadow as a function of time.
\subsection{Static observer}
In this section we focus on the description of the shadow as determined by a static observer. We assume that there are light sources distributed everywhere in the space-time except in the region between the observer and the matter distribution.
For $r>r_{+}$, we define an orthonormal tetrad for a static observer as
\begin{equation}
    e_0=\frac{1}{\sqrt{1-\frac{2M(r)}{r}}}\frac{\partial}{\partial T} \ , \qquad
    e_1=\sqrt{1-\frac{2M(r)}{r}}\frac{\partial}{\partial r}+\frac{\sqrt{2M(r)/r}}{\sqrt{1-\frac{2M(r)}{r}}}\frac{\partial}{\partial T} \ ,
    \label{e1}
\end{equation}
and
\begin{equation}
    e_2=\frac{1}{r}\frac{\partial}{\partial\theta} \ , \qquad     e_3=\frac{1}{r\sin{\theta}}\frac{\partial}{\partial\varphi} \ .
    \label{e2}
\end{equation}
The tangent vectors of null geodesics that arrive to a static observer can be written in terms the basis associated to the observer via the angle $\alpha$ that the vector makes with the corresponding radial vector as
\begin{equation}
        \Dot{T}\frac{\partial}{\partial T}+\Dot{r}\frac{\partial}{\partial r}+\Dot{\varphi}\frac{\partial}{\partial\varphi}=e_0+\cos{\alpha}e_1-\sin{\alpha}e_3 \ .
    \label{eq:chialp}
\end{equation}
where the dot denotes the derivative with respect to an affine parameter.
Equating the radial basis \eqref{e1} and the radial component of the four velocity
of equation~\eqref{eq:chialp} implies 
\begin{equation}
    \begin{split}
        \Dot{r}&=\sqrt{1-\frac{2M(r)}{r}}\cos{\alpha} 
         \ .
    \end{split}
\end{equation}
With the vector $e_2$ given by \eqref{e2} and the $\varphi$ component of \eqref{eq:chialp} one gets
\begin{equation}
    \begin{split}
        \Dot{\varphi}&=\frac{-\sin{\alpha}}{r} \ .
    \end{split}
\end{equation}
With the last set of equations one can compute $\frac{dr}{d\varphi}=\frac{ \dot r}{\dot \varphi}$ so that one arrives to
\begin{equation}
    \begin{split}
     \left(   \frac{dr}{d\varphi} \right)^2=\frac{(1-2M(r)/r)\cos^2{\alpha}}{\sin^2{\alpha}}\ .
    \end{split}
    \label{rphichialpha}
\end{equation}
 Using the equation~\eqref{rphirm} for null geodesics at $r=r_m$ on the previous equation, one obtains after some simplifications
\begin{equation}
    \frac{r^2(r_m-2M(r_m))}{r_m^3}-1+\frac{2M(r)}{r}=\frac{(1-2M(r)/r)\cos^2{\alpha}}{\sin^2{\alpha}} \ .
    \label{eq:alphastat}
\end{equation}
Solving \eqref{eq:alphastat} for $\alpha$, we finally get the angular size of the shadow as
\begin{equation}
    \sin{\alpha}=\sqrt{\mathcal{R}(r,r_m)}
    \label{alphas}
\end{equation}
where 
\begin{equation}\label{eq:rcal}
 \mathcal{R}(r,r_m)=\frac{r_m^3(r-2M(r))}{r^3(r_m-2M(r_m))}   \ .
\end{equation}

\subsection{Free falling observer}
As for the static observer, it is possible to define an orthonormal tetrad for a radially infalling observer with $\varepsilon>0$ as:
\begin{equation}
    \Tilde{e}_0=\frac{\varepsilon-\sqrt{2M(r)/r}\sqrt{\varepsilon^2-1+2M(r)/r}}{1-2M(r)/r}\frac{\partial}{\partial T}-\sqrt{\varepsilon^2-1+2M(r)/r}\frac{\partial}{\partial r} \ ,
    \label{e0s}
\end{equation}
\begin{equation}
    \Tilde{e}_1=\varepsilon\frac{\partial}{\partial r}+\frac{\varepsilon\sqrt{2M(r)/r}-\sqrt{\varepsilon^2-1+2M(r)/r}}{1-2M(r)/r}\frac{\partial}{\partial T}
    \label{e1s}
\end{equation}
and
\begin{equation}
    \Tilde{e}_2=\frac{1}{r}\frac{\partial}{\partial\theta}
    \label{e2s} \ , \qquad     \Tilde{e}_3=\frac{1}{r\sin{\theta}}\frac{\partial}{\partial\varphi} \ .
\end{equation}
Considering the angle $\tilde \alpha$, that null rays made in the system of the free falling observer one gets
\begin{equation}
    \begin{split}
        \Dot{T}\frac{\partial}{\partial T}+\Dot{r}\frac{\partial}{\partial r}+\Dot{\varphi}\frac{\partial}{\partial\varphi}=\Tilde{e}_0+\cos{\Tilde{\alpha}}\Tilde{e}_1-\sin{\Tilde{\alpha}}\Tilde{e}_3 \ .
    \end{split}
    \label{eq:chialptilde}
\end{equation}
From this expression and Eq. \eqref{rtaus} one may write $\Dot{r}$ and $\Dot{\varphi}$ in terms of $\tilde \alpha$ as
\begin{equation}
\Dot{r}=\varepsilon\cos{\Tilde{\alpha}}-\sqrt{\varepsilon^2-1+2M(r)/r} \ ,  \qquad  \Dot{\varphi}
        =-\frac{\sin{\Tilde{\alpha}}}{r} \ .
\end{equation}
Taking the quotient, one gets the trajectory in terms of the angular size.
\begin{equation}
    \begin{split}
        \frac{dr}{d\varphi}=\frac{r\left(\varepsilon\cos{\Tilde{\alpha}}-\sqrt{\varepsilon^2-1+2M(r)/r}\right)}{-\sin{\Tilde{\alpha}}} \ .
    \end{split}
    \label{eq:rphichialphati}
\end{equation}
Now, by using~\eqref{rphirm} and~\eqref{eq:rphichialphati} for a free falling observer
\begin{equation}
    \frac{r^2(r_m-2M(r_m))}{r_m^3}-1+\frac{2M(r)}{r}=\frac{\left(\varepsilon\cos{\Tilde{\alpha}}-\sqrt{\varepsilon^2-1+2M(r)/r}\right)^2}{\sin^2{\Tilde{\alpha}}}
    \label{eq:alphamov}
\end{equation}
the corresponding expression for the angular size $\tilde \alpha$, can be obtain by solving \eqref{eq:alphamov}
\begin{equation}
    \sin{\Tilde{\alpha}}=\frac{\sqrt{1-2M(r)/r}\sqrt{\mathcal{R}(r,r_m)}}{\varepsilon\pm\sqrt{\varepsilon^2-1+2M(r)/r}\sqrt{1-\mathcal{R}(r,r_m)}} \ .
    \label{alpham}
\end{equation}
The upper sign in the denominator of equation~\eqref{alpham} applies when $dr/d\varphi>0$, and the lower sign applies when $dr/d\varphi<0$.
Equations~\eqref{alphas} and~\eqref{alpham} establish a relation between the angle of the dark region (the shadow) and the observer's radius, depending on the minimum radius $r_m$ of the collapsing star. 
Equation~\eqref{alphas} 
is valid for 
$r>r_+$. 
On the other hand, equation~\eqref{alpham} 
is valid for $r>0$ if $\varepsilon\geq1$.

\section{Shadow of the collapsing star}
\label{sec:shadow_colapse}
\begin{figure}
\centering
\subfloat[$q_*=0.0$]{\includegraphics[scale=0.24]{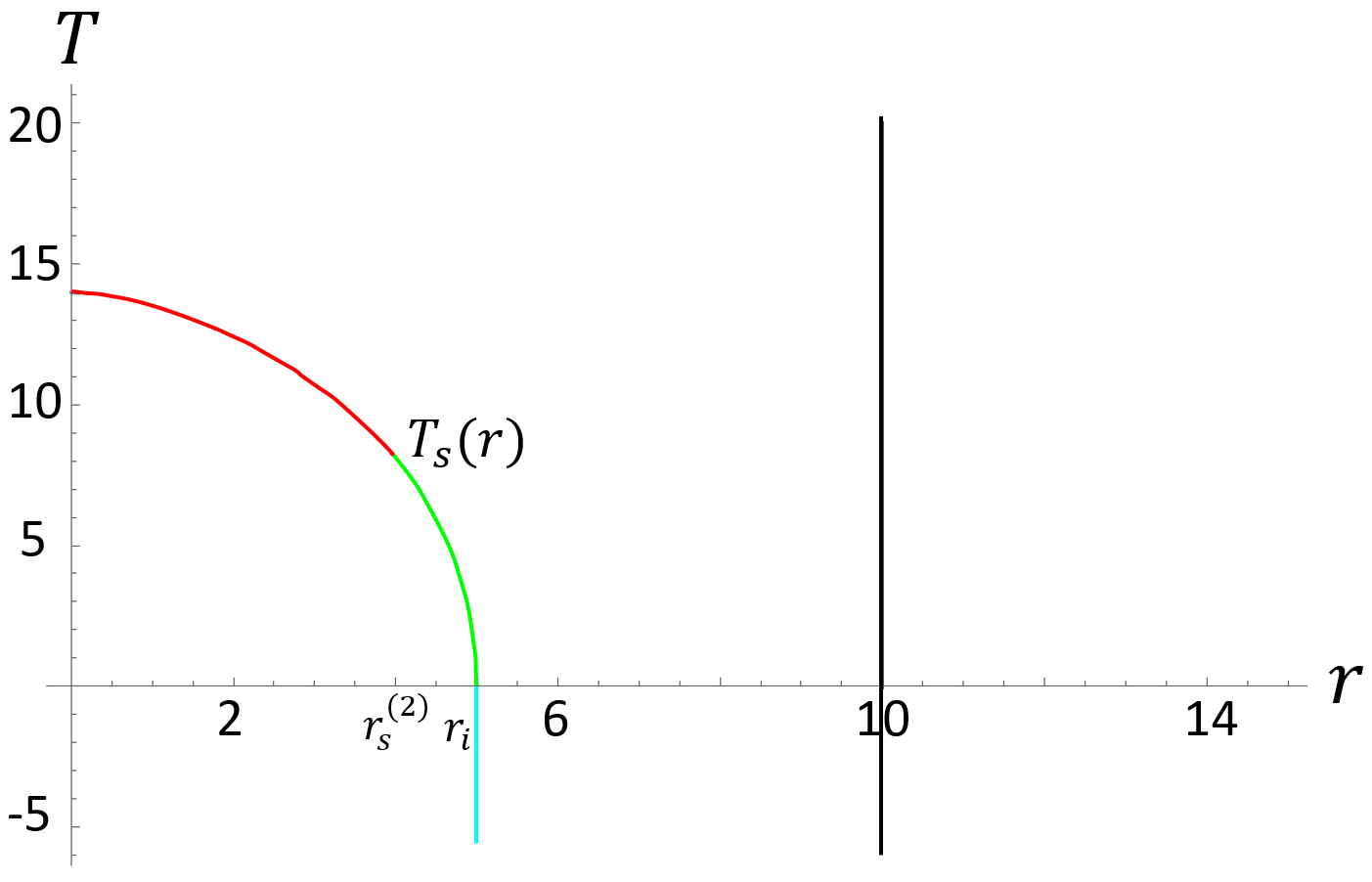}} 
\subfloat[$q_*=0.8967$]{\includegraphics[scale=0.24]{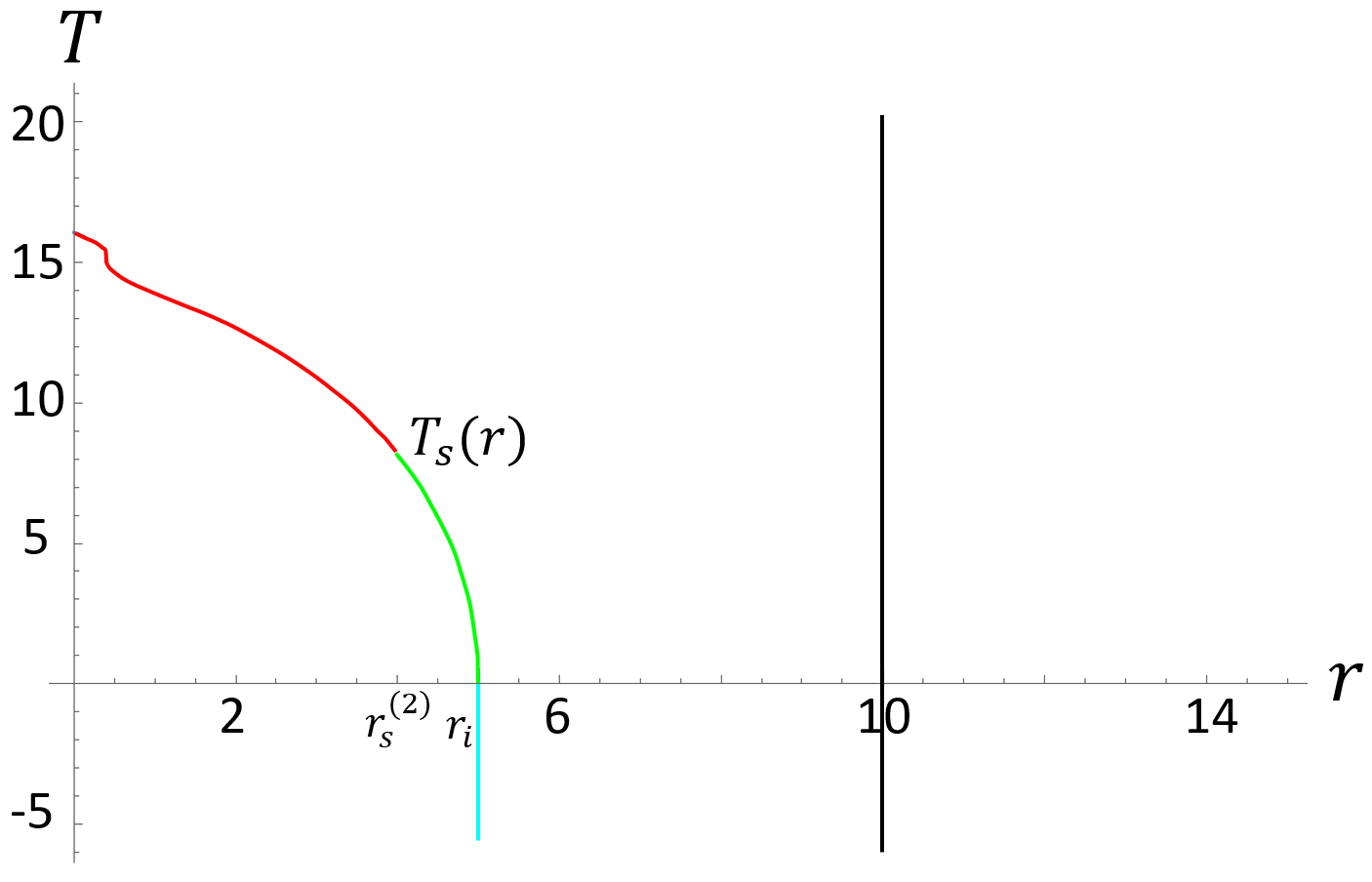}}
\subfloat[$q_*=1.0582$]{\includegraphics[scale=0.24]{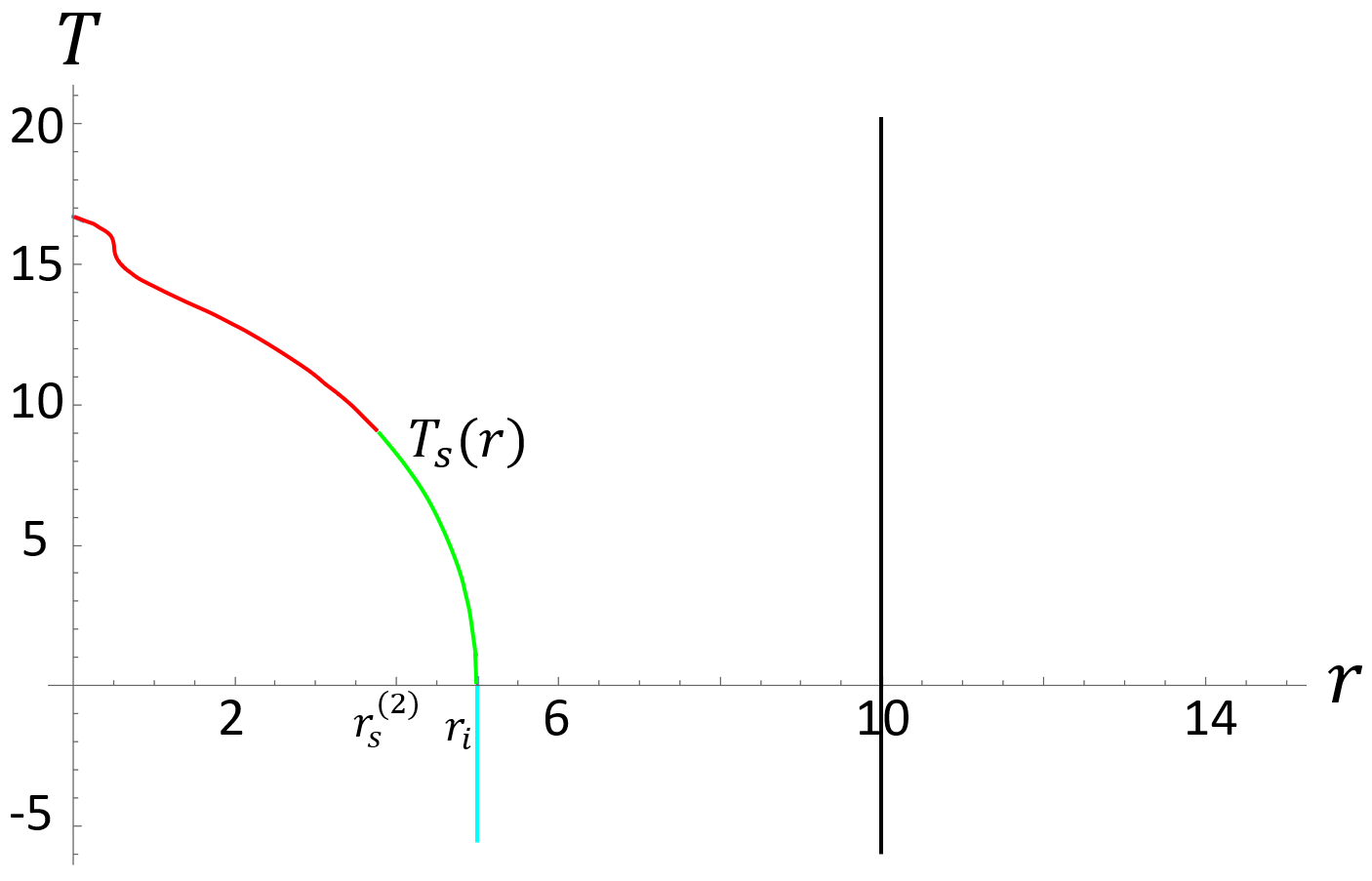}} 
\caption{Space time diagram for the collapse of spherical dust for different values of $q_*$.
The curve $T_s(r)$ denotes the trajectory of the surface of the star $r_s$ during the collapse.
The static observer is at $r_o=10$.
}
\label{grcolesf}
\end{figure}

Following \cite{Schneider:2018hge} we make a description of the development in time of the shadow, as given by the angular size, in three phases. In the first phase from $T_s <0$ to $T_s = 0$, the star has a constant radius $r_s=r_i$. In the second phase that last from $T_s=0$ to $T_s=T_s^{(2)}$ the star collapses to a value $r_s=r_s^{(2)}$. In the third phase from $T_s=T_s^{(2)}$ to  
$T_s=T_s^{\rm coll}$ the star completes the collapse to $r_s=0$.
At the beginning of the collapse at
$T_s=0$ the surface of the star is initially at $r_i$, thus 
the energy of each particle on the surface of the star is given by
$\varepsilon^2=1-2M(r_i)/r_i$.
As the surface contracts from an initial radius to a radius 
$r_s$ the time elapsed 
can be obtained from the quotient of Eq.~\eqref{ttaus} and Eq.~\eqref{rtaus} yielding 
\begin{equation}
    T_s=\int^{r_i}_{r_s}\frac{\sqrt{r^3}\sqrt{r_i-2M(r_i)}}{(r-2M(r))\sqrt{2M(r)r_i-2M(r_i)r}}-\frac{\sqrt{2M(r)r}}{r-2M(r)}dr
    \label{tsrs}
\end{equation}
Figure~\ref{grcolesf} shows the time $T_s$ against the radius of the surface of the star for some values of $q_*=q/M_{0}$. Remind that $q_* = 0$ corresponds to the Schwarzschild black hole and $q_*=1.0583$ corresponds to an extreme Hayward. The position of the static observer is also shown for reference.
Figure~\ref{grcolesf} also shows the three different stages of the description of the shadow. As the value of $q_*$ increases the time that takes the boundary of the star to get $r_s=0$ increases. Furthermore for larger values of $q_*$ the curve $T_s(r)$ presents a change in its slope close to the origin due to the repulsive nature of the core in the Hayward spacetime.

Substituting the position of the surface of the star $r=r_s$, a maximum angular amplitude
 $\Tilde{\alpha}=\pi/2$ and $\varepsilon^2=1-2M(r_i)/r_i$ in the equality given by Eq.~\eqref{eq:alphamov}, one obtains
\begin{equation}
    \frac{r_m^3}{r_m-2M(r_m)}=\frac{r_ir_s^2}{r_i-2M(r_i)} \ .
    \label{rmrs}
\end{equation}
As $r_s$ evolves during the collapse, the radius $r_m$ 
decreases up to the limiting value $r_{mp}$ given in~\eqref{rm}. 
The value of $r_s$, for which $r_m=r_{mp}$ is reached, will be denoted as $r_s^{(2)}$ and marks the end of the second phase. In table~\ref{Importantradius} the values for the horizon radius $r_{+}$, the photo sphere radius $r_{mp}$ and the radius for the end of second phase $r_s^{(2)}$ for representative values of $q_*$ are shown.
\begin{table}
\begin{center}
\begin{tabular}{|l|l|l|l|}
\hline
$q_*$ & $r_{+}$ & $r_{mp}$ & $r_s^{(2)}$ \\ \hline \hline
0.0 & 2.0 & 3.0 & 4.02492 \\ \hline
0.8967 & 1.76981 & 2.8151 & 3.91263\\ \hline
1.0582 & 1.33985 & 2.6524 & 3.82057\\ \hline
\end{tabular}
\caption{The Schwarzschild case corresponds to $q_*=0$ and the limit value of $q_*=1.0582$ corresponds to an extreme Hayward BH. As the value of $q_*$ increases the other radii decrease.}
\label{Importantradius}
\end{center}
\end{table}

The description of the shadow will depend on the relative position between the observer $r_o$, and  the surface of the star. Thus the static case will be divided in two cases: when the observer is always outside the star $r_o>r_i$, and when for a moment, the observer is inside the star $r_s^{(2)}<r_o\leq r_i$.

\subsection{Static observer outside the star 
}
As in the previous case the description of the shadow will be divided in three stages. 
In the first stage, before the collapse starts, 
the static observer describes a shadow with a constant angular size given by Eq.~\eqref{alphas}, where we have set $r_m=r_i$ and $r_i=r_o$ that corresponds to the position of the observer. %
\begin{equation}
    \sin{\alpha}=\sqrt{\frac{r_i^3(r_o-2M(r_o))}{r_o^3(r_i-2M(r_i))}} \ .
    \label{alpha1}
\end{equation}
The beginning of the collapse marks the end of the first stage. %
Integrating Eq.~\eqref{ttray} from $r=r_i$ to $r=r_o$, we obtain $T_o^{(1)}$, that is the time elapsed for a photon to travel from $r_m=r_i$ to $r_o$.

\begin{equation}
    T_o^{(1)}=\int^{r_o}_{r_i}\frac{\sqrt{2M(r)r}}{r-2M(r)}+\frac{\sqrt{r^5(r_i-2M(r_i))}}{(r-2M(r))\sqrt{(r_i-2M(r_i))r^3-(r-2M(r))r_i^3}}dr \ .
\end{equation}

During the second stage, as the collapse is taking place, the shadow will be modeled based on the minimum radius through which the photons pass $r_m$. 
This minimum can be determined using the radius of the star's surface $r_s$ for a given value of $r_i$ as given in Eq.~\eqref{rmrs}.
We can use this expression for $r_m=r_m(r_s)$  in Eq. \eqref{eq:rcal} to get
\begin{equation}
    \mathcal{R}(r,r_m(r_s)) = \frac{r_ir_s^2(r-2M(r)}{r^3(r_i-2M(r_i))} \ .
\end{equation}
For the observer at $r=r_0$,
the angular size of the shadow as a function of the surface of the star is given by means of Eq. \eqref{alphas} as
\begin{equation}   \sin{\alpha}=\sqrt{\frac{r_ir_s^2(r_o-2M(r_o))}{r_o^3(r_i-2M(r_i))}} \ .
    \label{alpha2}
\end{equation}
From equation~\eqref{ttray} we can determine the time that photons take to travel from $r_s$ to the observer's location at $r_o$. 
\begin{equation}
    T_o-T_s=\int^{r_o}_{r_s}\frac{\sqrt{2M(r)r}}{r-2M(r)}+\frac{\sqrt{r^5(r_i-2M(r_i))}}{(r-2M(r))\sqrt{(r_i-2M(r_i))r^3-(r-2M(r))r_ir_s^2}}dr \ ,
    \label{T0-2}
\end{equation}
where $T_s$ given by equation~\eqref{tsrs}. Therefore, with the star's radius, one can find an implicit relation between the shadow angular size $\alpha$ and the time the photons that leave the surface of the star
reach the position of the static observer $T_o$.

\begin{figure}
\subfloat[$q_*=0.0$, $r_o=10$]{\includegraphics[scale=0.42]{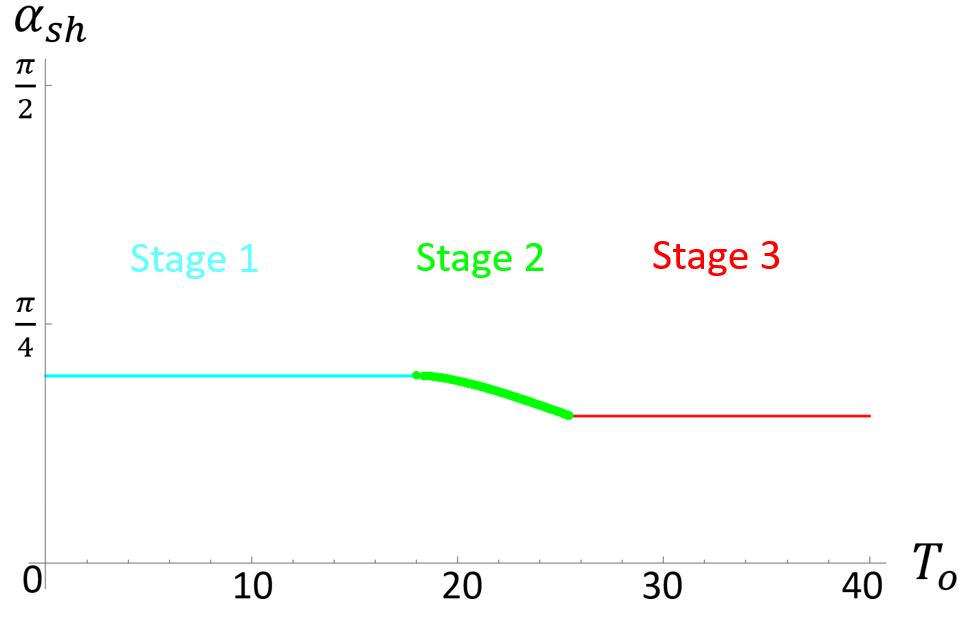} \includegraphics[scale=0.28]{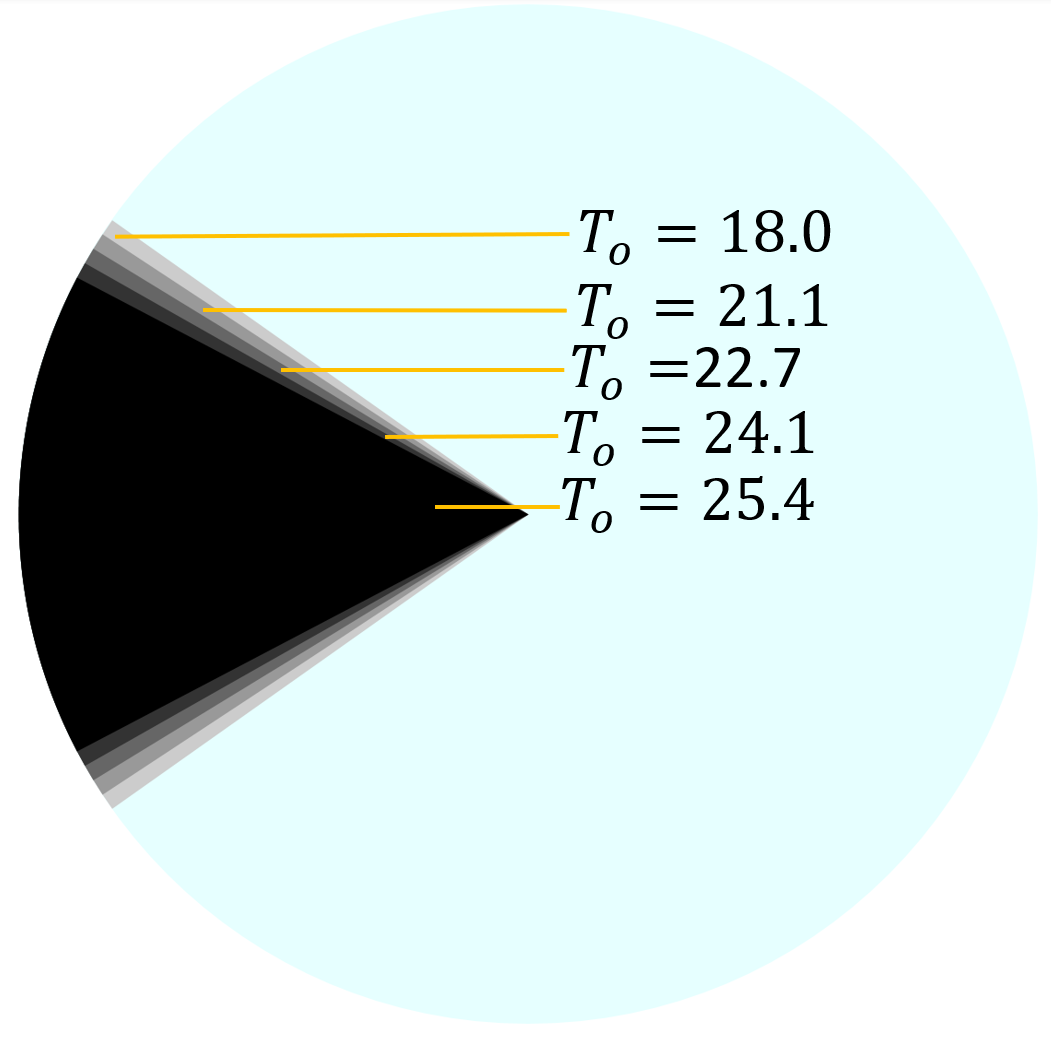}}\\
\subfloat[$q_*=0.8967$, $r_o=10$]{\includegraphics[scale=0.44]{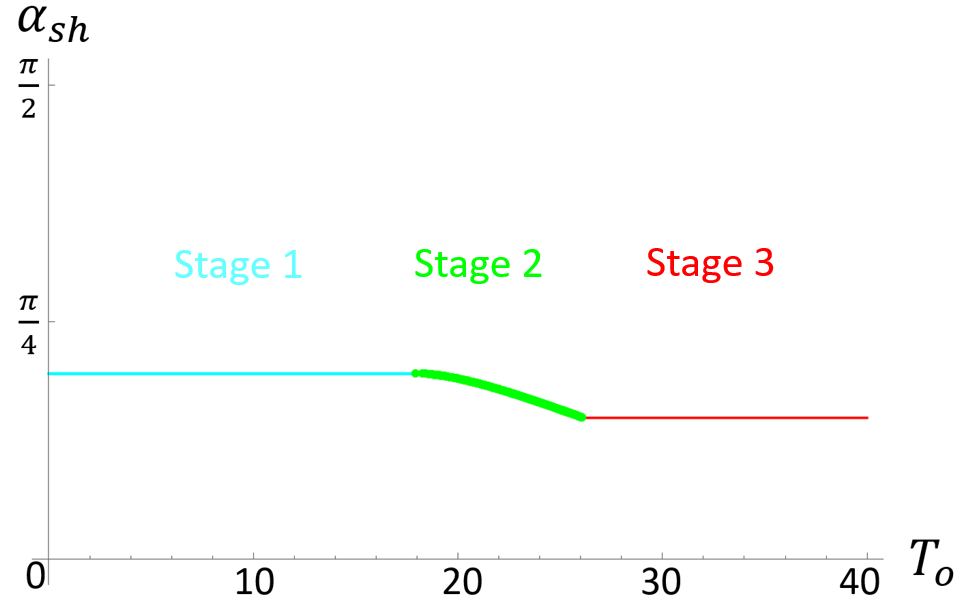}\includegraphics[scale=0.3]{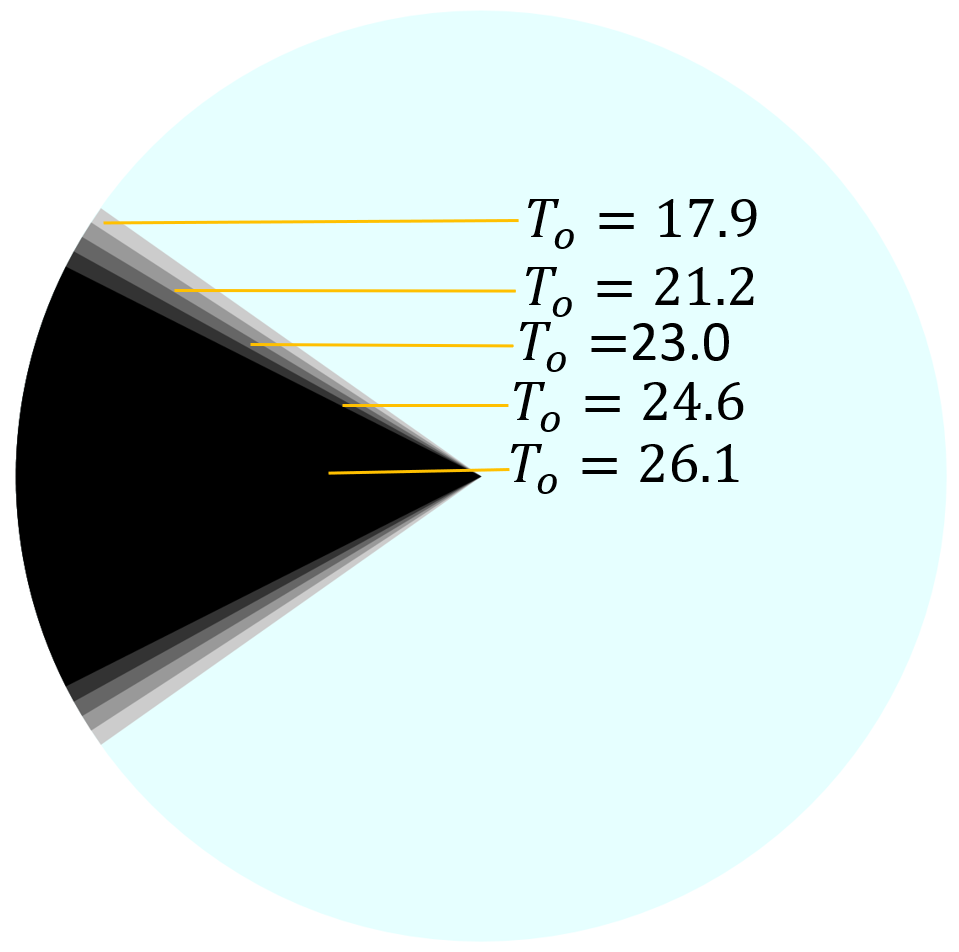}}\\
\subfloat[$q_*=1.0582$, $r_o=10$]{\includegraphics[scale=0.44]{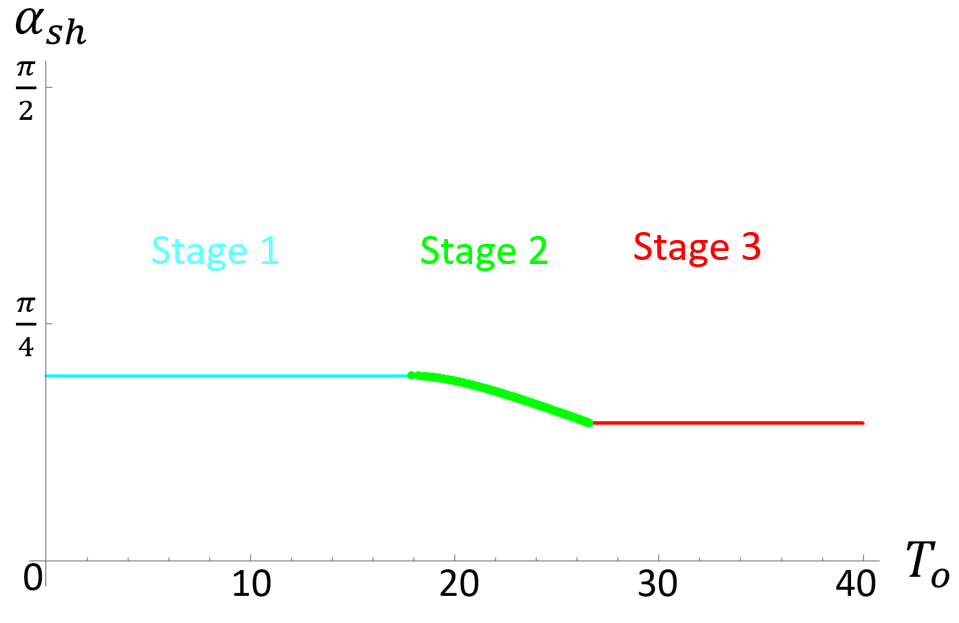}\includegraphics[scale=0.3]{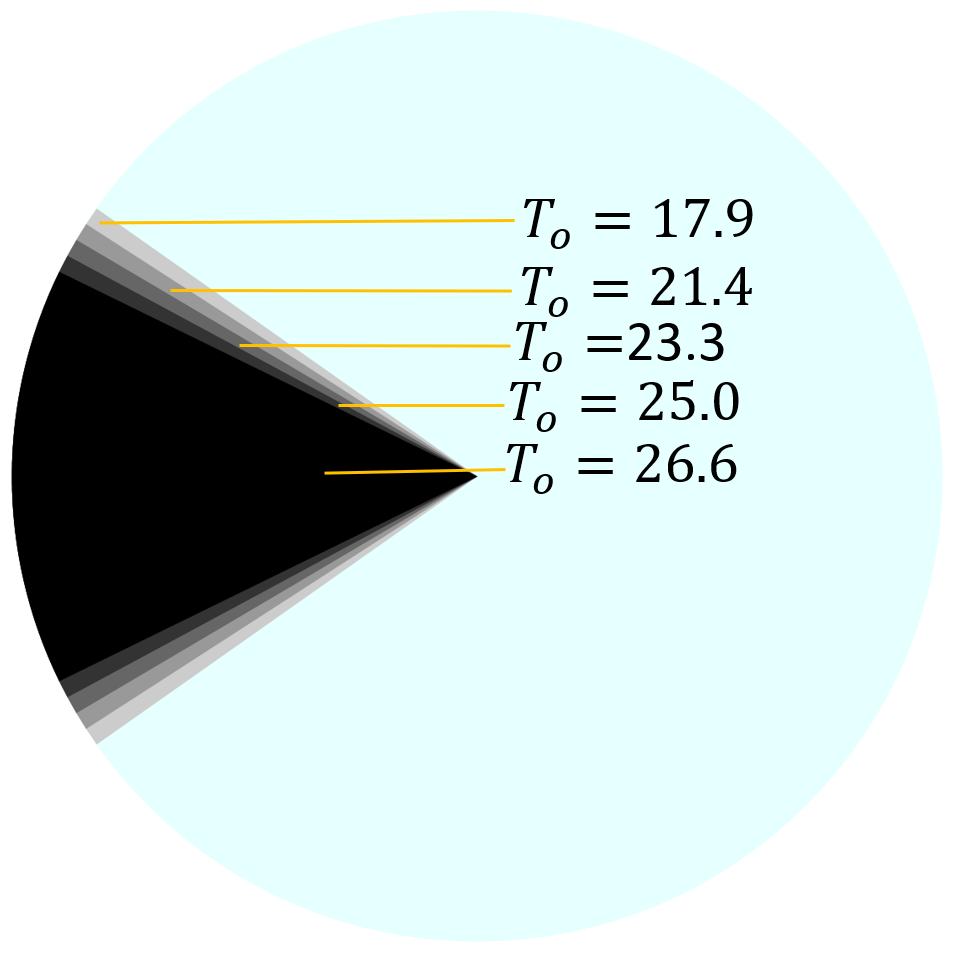}}
\caption{(Left) Angular size of the shadow $\alpha_{sh}$ as a function of time $T_o$ measured by the static observer at $r_o>r_i$ for different values of $q_*$. The color in the plot indicates the three different phases of the description. The first phase is the cyan part of the curve (equation \eqref{alpha1}), the second stage is the green part (equation \eqref{alpha2}) and the red color corresponds to the third and final phase (equation \eqref{alpha3}). 
(Right) Angular region of the shadow at different times. The black hole is to the left of the picture
The initial radius is $r_i=5$.}
\label{gr3fases}
\end{figure}

The second stage ends when the minimum radius coincides with the limit value $r_m=r_{mp}$, we label the star's radius at the end of this stage as $r_s=r_s^{(2)}$. 
The elapsed time during this stage is given by
\begin{equation}
    \begin{split}
        T_o^{(2)}&=T_s(r_s^{(2)})+\int^{r_o}_{r_s^{(2)}}\frac{\sqrt{2M(r)r}}{r-2M(r)}dr\\
        &+\int^{r_o}_{r_s^{(2)}}\frac{\sqrt{r^5(r_i-2M(r_i))}}{(r-2M(r))\sqrt{(r_i-2M(r_i))r^3-(r-2M(r))r_ir_s^{(2)2}}}dr \ .
    \end{split}
\end{equation}
The third stage is characterized because 
the minimum radius acquires a constant value $r_{\rm m}=r_{\rm mp}$. During this last stage, the photons coming from the source that reach the observer can graze, at most, the photosphere.
Furthermore, during this stage
the angular size remains constant, and is given by equation~\eqref{alphas} with $r=r_o$ and $r_m=r_{mp}$.
\begin{equation}
\sin{\alpha}=\sqrt{\frac{r_{mp}^3(r_o-2M(r_o))}{r_o^3(r_{mp}-2M(r_{mp}))}} \ .
    \label{alpha3}
\end{equation}
The first row of panels of Figure~\ref{gr3fases} displays the Schwarschild case ($q*=0$) for the angular size of the shadow for an static observer. In the left panel it it plot the angular size as a function of time for the observer (which coincides with the coordinate time). In the right panel the observer's sky is represented. 
In this representation we have considered that the black hole is at the left and the darker region represent the evolution of the shadow as a function of time. The gray scale color code matches the value of $\alpha$. The darker the region the higher the value of $\alpha$. 
The panels in the second and third rows represent the shadow angular size for spacetimes with representative values of the Hayward parameter. The extreme Hayward black hole corresponds to the value $q_m/M_0=1.05827$.

As a summary of this scenario we have the following:
when the static observer is located outside the initial radius of the star $r_i$, the determination of the shadow is divided into three stages: the first stage corresponds to photons that leave the source before the collapse begins and reach the observer, the second stage spans from the start of the collapse until the minimum radius at which photons can orbit ($r_m$) becomes equal to the value of the photosphere ($r_m = r_{mp}$) and the third stage covers the rest of the collapse. In this last stage, the radius of the star ($r_s$) continues decreasing but becomes irrelevant for the shadow determination, since the minimum radius remains fixed at $r_{mp}$. 
For the static observer, the only change in the angular size of the shadow occurs during the second stage of the collapse when the minimum radius $r_m$ changes from $r_i$ to $r_{mp}$. This makes sense because the photons marking the edge of the shadow follow circular trajectories that depend on the size of the star until they reach that minimum limit.
For the Schwarzschild case, for different initial observer radii, the closer the observer is to the star, the larger the shadow angle is. Additionally, the start of each stage begins earlier for smaller observer radii, as it takes less time for photons to travel from the star's surface to the observer. For values of $q_*>0$, a higher value of the Hayward charge parameter leads to the second stage ending at a later $T_o$, as the photosphere's radii value is smaller, causing photons to reach a smaller radius in their circular orbits. Since the second stage ends at a later times and $r_{mp}$ is smaller for small values of $q*$, it results in a smaller angular size in the last stage.

\subsection{Static observer located initially inside the star.}

In contrast to the previous case, the description of the shadow can only begin when the star's radius is smaller than the observer's radius since our model considers a opaque star. We also assume that $r_s^{(2)}<r_o$. The equations that define the angular size of the shadow correspond to Eq.~\eqref{alpha2} and Eq. ~\eqref{T0-2} in stage 2  Eq.~\eqref{alpha3} in stage 3.

\begin{figure}
\centering
\subfloat[$q_*=0.0$]{\includegraphics[scale=0.42]{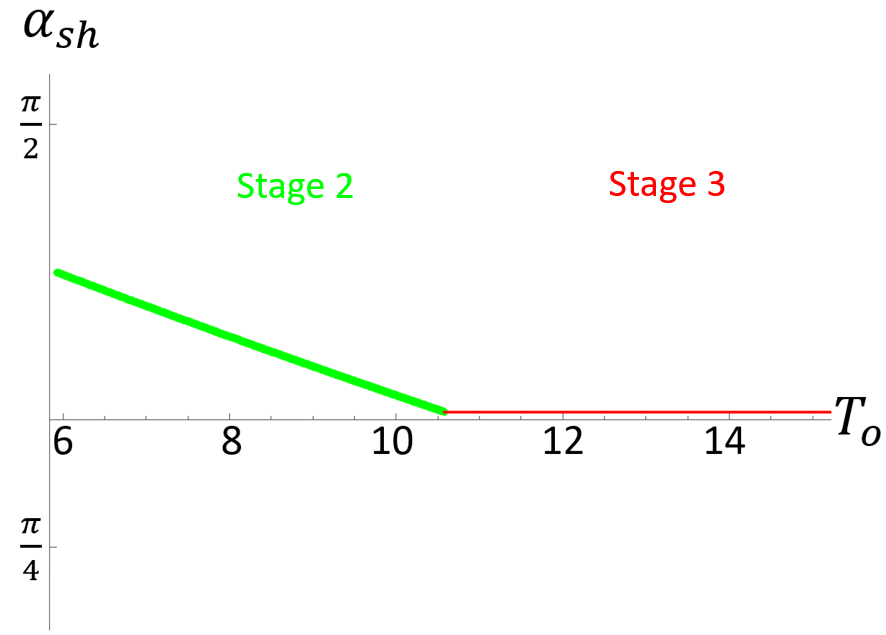}
\includegraphics[scale=0.25]{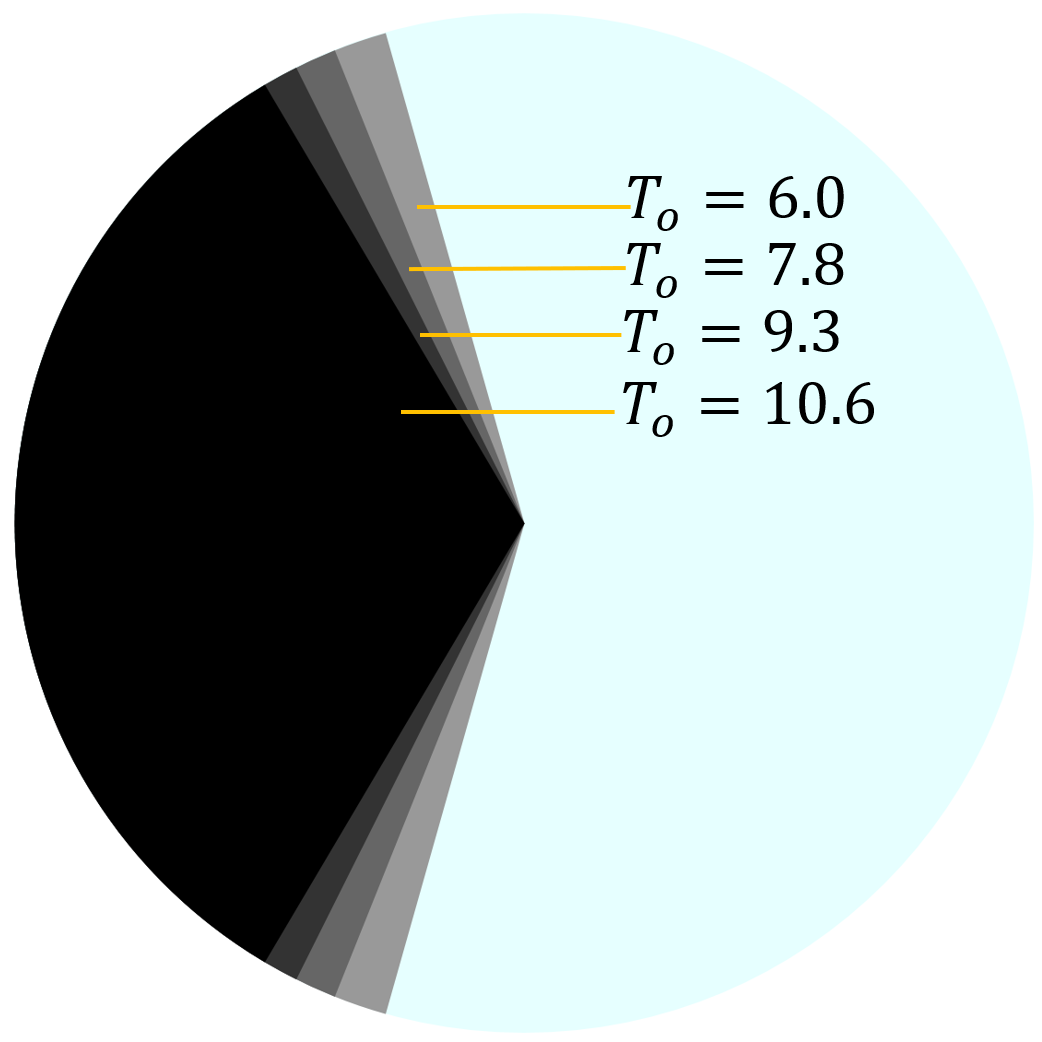}} \\
\subfloat[$q_*=0.8967$]{\includegraphics[scale=0.41]{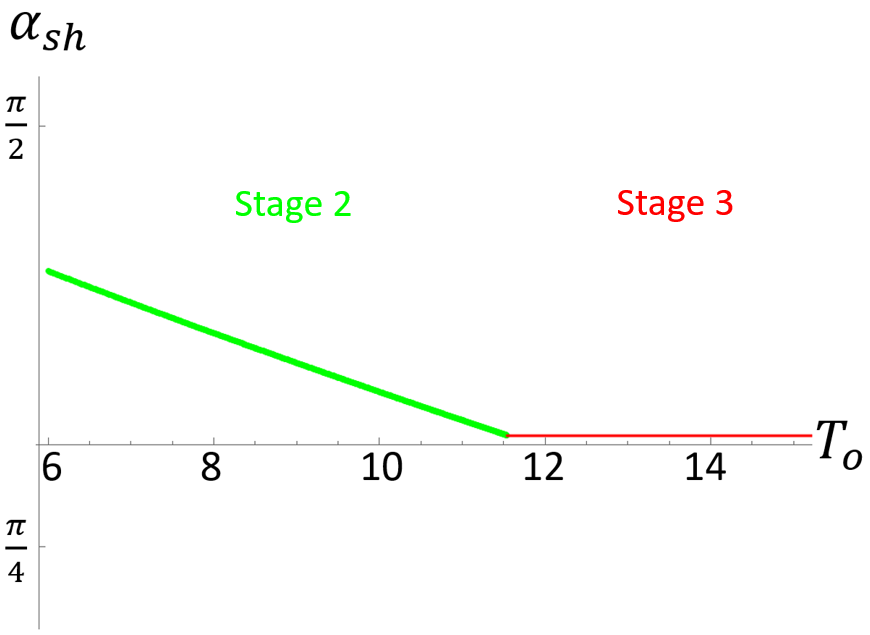}
\includegraphics[scale=0.25]{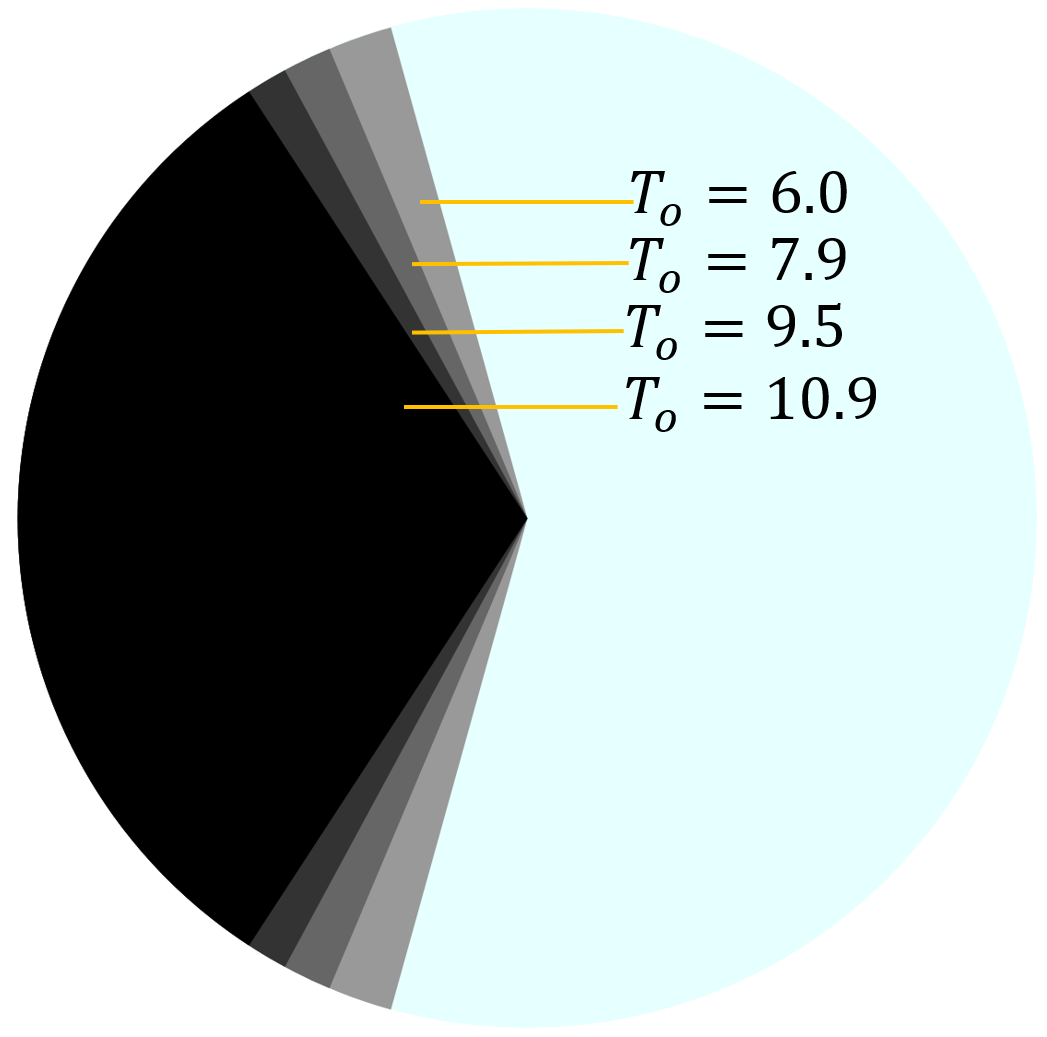}
}\\
\subfloat[$q_*=1.0582$]{\includegraphics[scale=0.4]{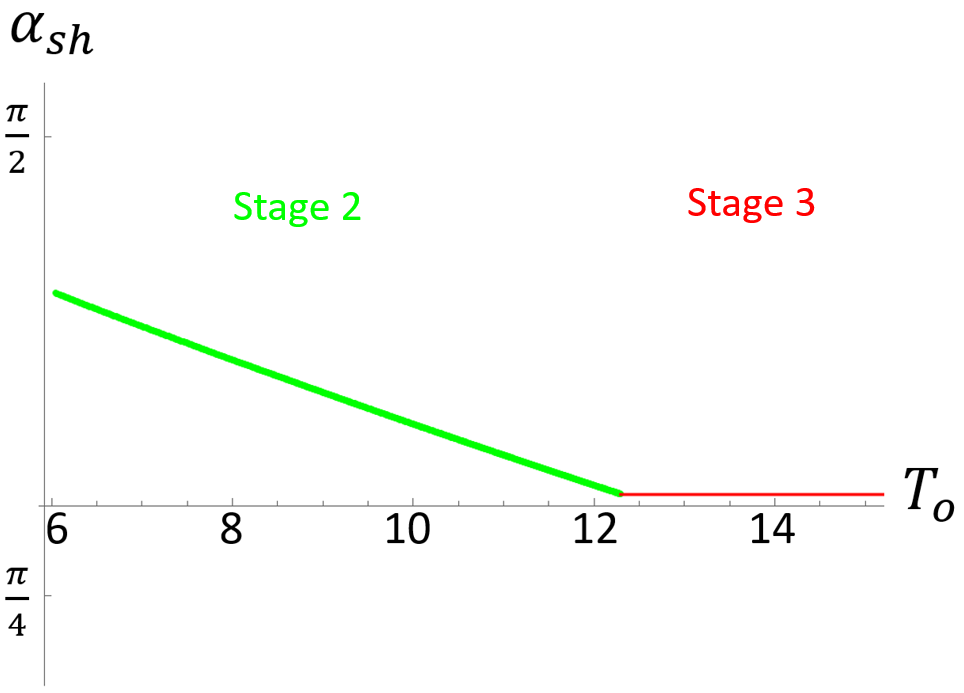}
\includegraphics[scale=0.32]{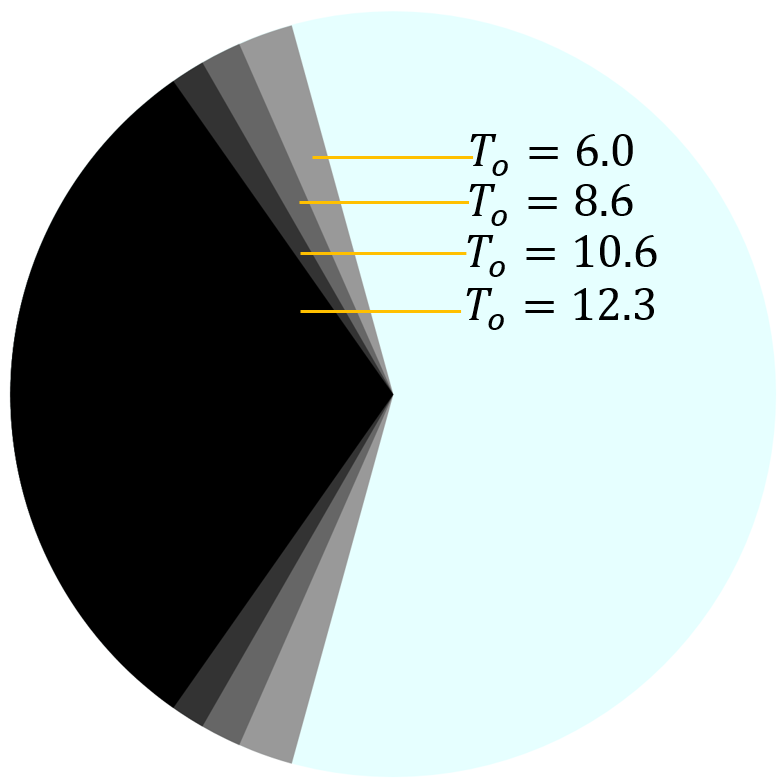}}
\caption{The shadow angular size $\alpha_{sh}$ as a function of $T_o$ determined by a static observer at $r_o=4.5$ for different values of $q_*$. }
\label{gr2fases}
\end{figure}

Figure~\ref{gr2fases} shows the evolution of $\alpha$ on time $T_o$ measured by the static observer at $r_o$ for different values of $q_*$.
For values of $q*$ near the extreme, the second stage last longer and the value of $\alpha$ is smaller
during the last stage for smaller values of $q*$. 

\subsection{Free falling observer 
}
For a free falling observer, the description of the shadow  will be also divided into three stages.
We shall consider that  
the initial radius of the star satisfies $r_i>r_{mp}$. 
We assume the observer is located initially at a radius $r_o^*$, large enough such that the observer is always outside the star.
The time elapsed for the observer from the beginning of the collapse $T_o=0$ up to an arbitrary time $T_o$ (when the observer is located at $r_o$ ) is calculated by taking the quotient of equations~\eqref{rtaus} and~\eqref{ttaus} and integrating
\begin{equation}    T_o=\int^{r_o^*}_{r_o}\frac{\varepsilon\sqrt{r^3}-\sqrt{2M(r)r}\sqrt{\varepsilon^2r-r+2M(r)}}{(r-2M(r))\sqrt{\varepsilon^2r-r+2M(r)}}dr
    \label{Tobs}
\end{equation}
The value of the energy $\varepsilon$ is determined by the initial velocity of the observer.

\begin{figure}
\centering
\subfloat[$q_*=0$]{\includegraphics[scale=0.22]{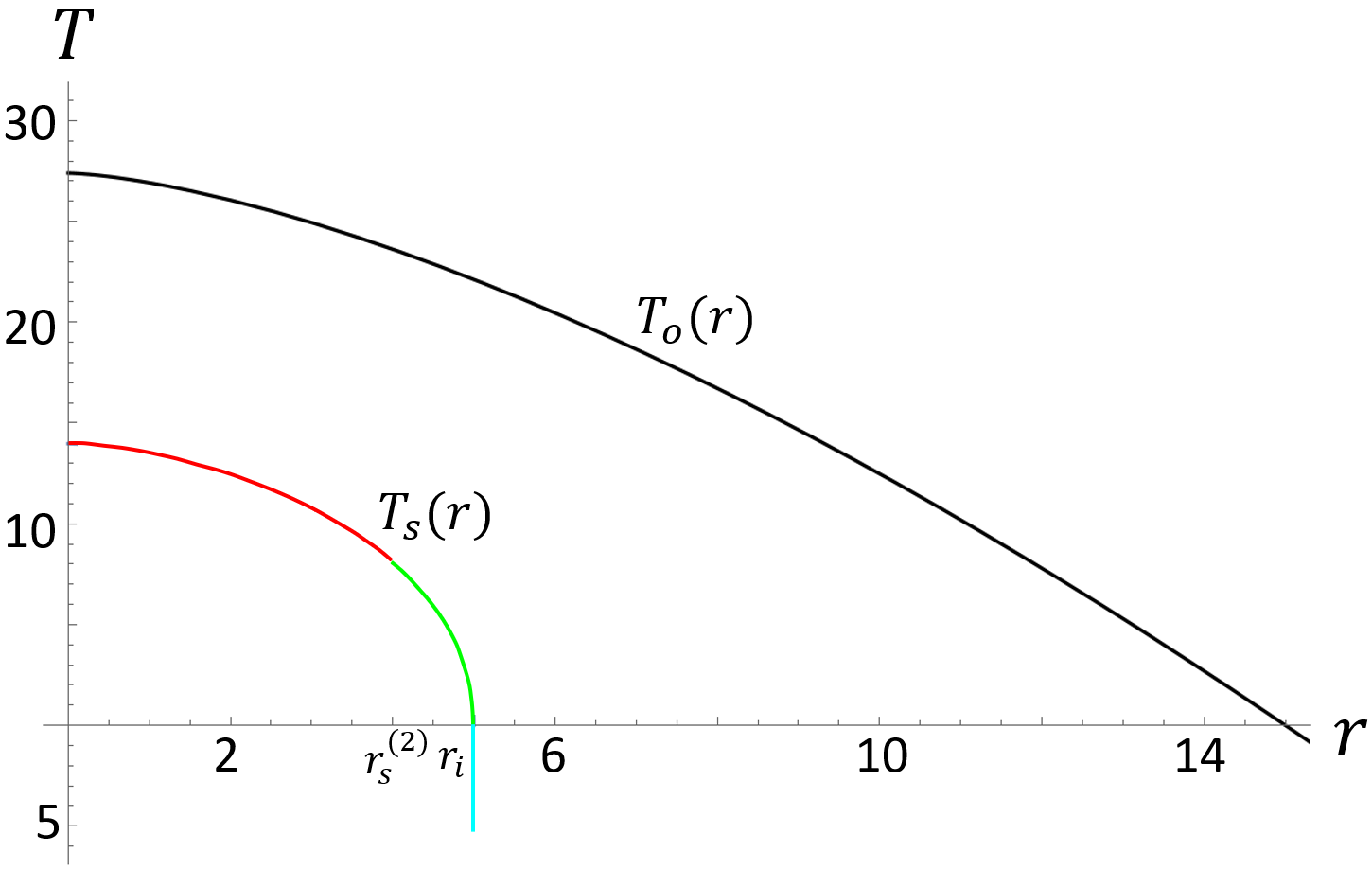}} 
\subfloat[$q_*=0.8967$]{\includegraphics[scale=0.22]{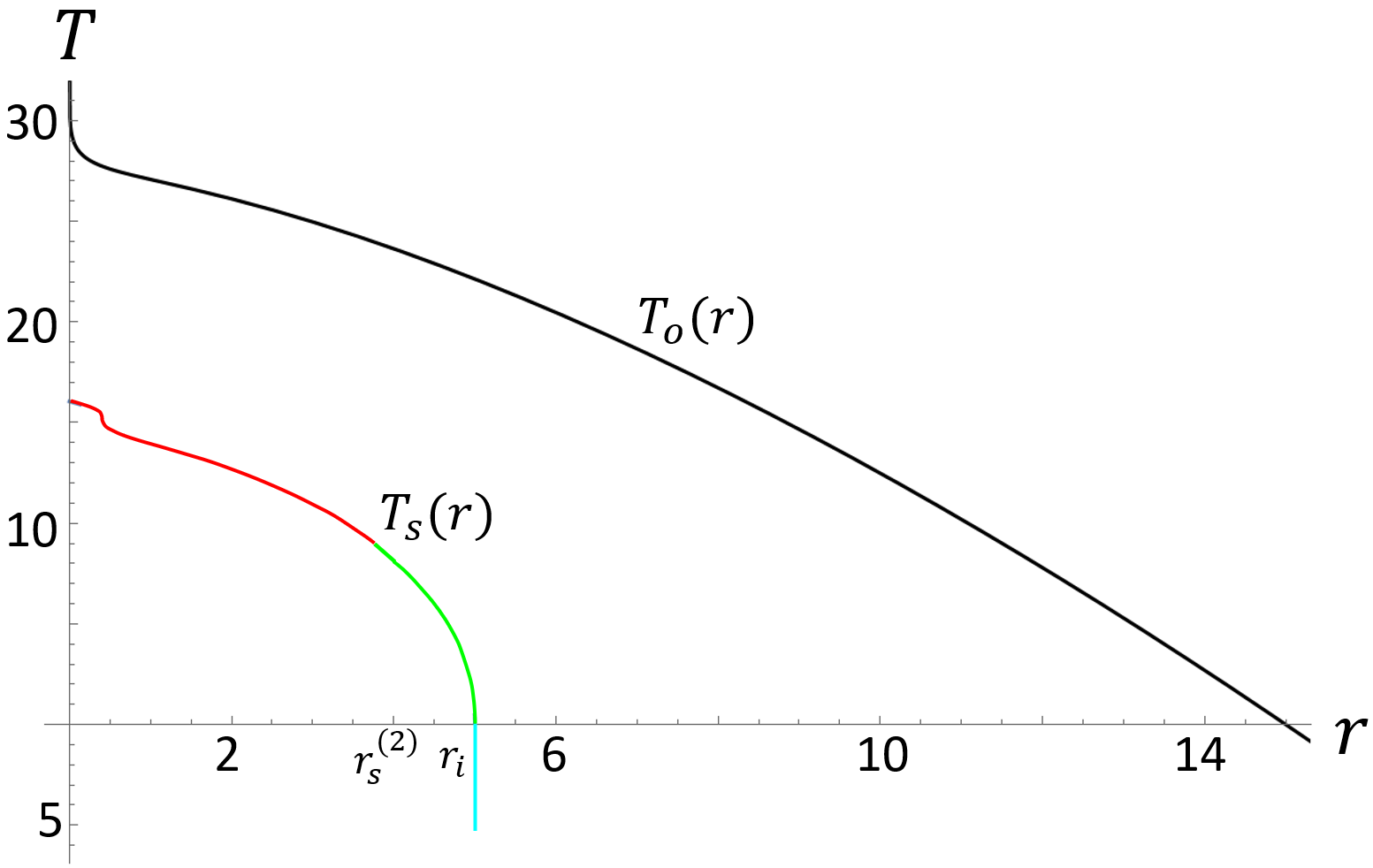}}
\subfloat[$q_*=1.0582$]{\includegraphics[scale=0.22]{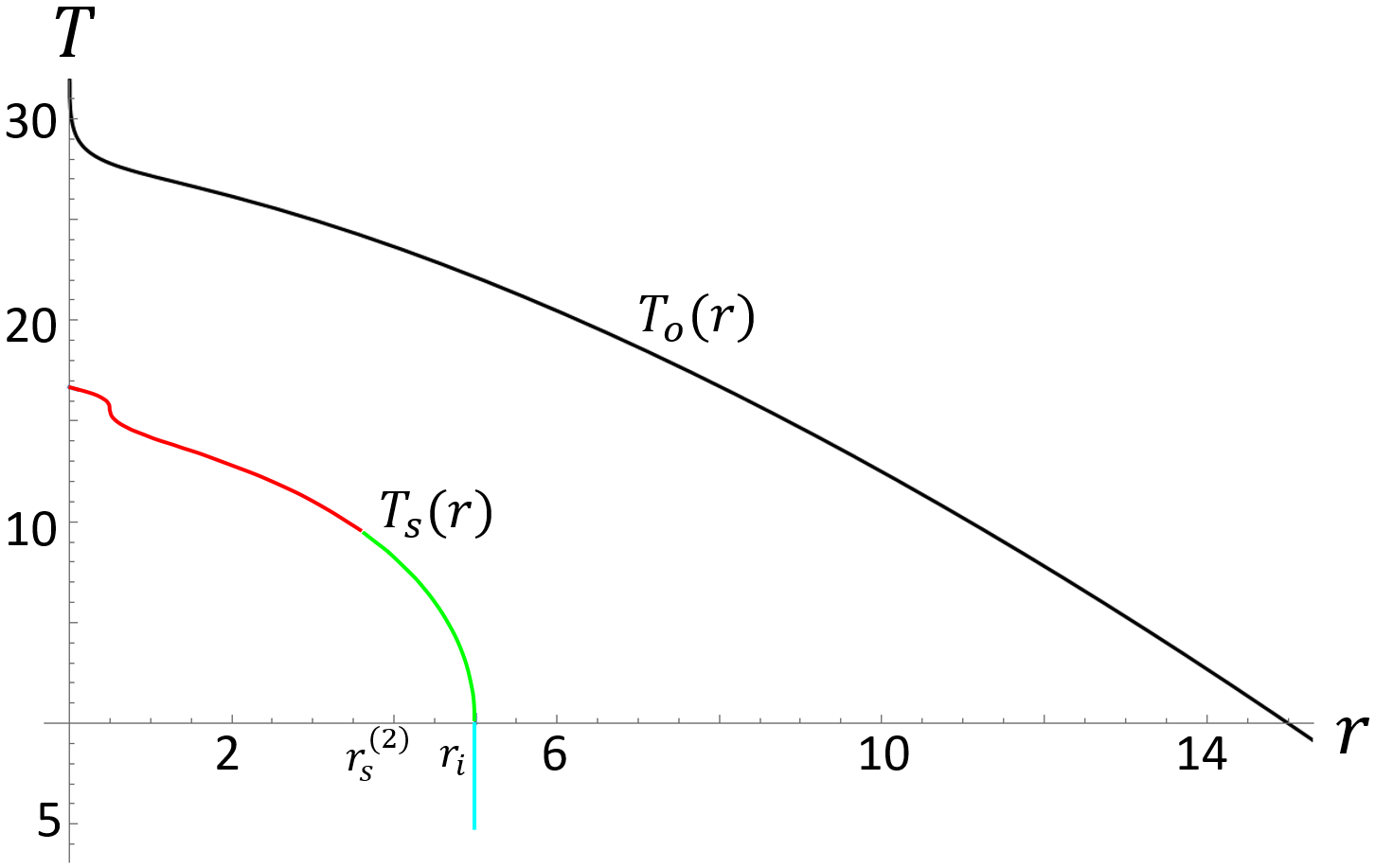}} 
\caption{Spacetime diagram of the surface of the collapsing star and the trajectory of a free falling observer for different values of $q_*$. The initial position of the observer is $r_o^*=15$.}
\label{grobscaida}
\end{figure}

In Figure~\ref{grobscaida} we plot the trajectory $T_s=T_s(r)$ of the surface of the collapsing star
alongside the trajectory of the falling observer $T_o=T_o(r)$ for some representative values of $q*$. 

For the first stage, that begins at an arbitrarily (negative) time ends when the collapse begins, the angular size shadow is determined by equation~\eqref{alpham} after taking $r=r_o$ and $r_m=r_i$.
\begin{equation}
    \sin{\Tilde{\alpha}}=\frac{\sqrt{1-2M(r_o)/r_o}\sqrt{\mathcal{R}(r_o,r_i)}}{\varepsilon+\sqrt{\varepsilon^2-1+2M(r_o)/r_o}\sqrt{1-\mathcal{R}(r_o,r_i)}}\ ,
\end{equation}
where $\mathcal{R}$ is given in Eq. \eqref{eq:rcal}.
Thus, the angular size of the shadow will change according to the position of the observer. 

In the second stage, where the radius of the surface of the star varies from $r_s=r_i$ to $r_s=r_s^{(2)}$, the equation~\eqref{T0-2} with $T_s$ given by the equation~\eqref{tsrs} and $T_o$ given by the equation~\eqref{Tobs} is: 
\begin{equation}
    \begin{split}
        \int^{r_o^*}_{r_o}\frac{\varepsilon\sqrt{r^3}-\sqrt{2M(r)r}\sqrt{\varepsilon^2r-r+2M(r)}}{(r-2M(r))\sqrt{\varepsilon^2r-r+2M(r)}}dr&=\int^{r_i}_{r_s}\frac{\sqrt{r^3}\sqrt{r_i-2M(r_i)}}{(r-2M(r))\sqrt{2M(r)r_i-2M(r_i)r}}dr+\int^{r_o}_{r_i}\frac{\sqrt{2M(r)r}}{r-2M(r)}dr\\
        &+\int^{r_o}_{r_s}\frac{\sqrt{r^5(r_i-2M(r_i))}}{(r-2M(r))\sqrt{(r_i-2M(r_i))r^3-(r-2M(r))r_ir_s^2}}dr
    \end{split}
\end{equation}
Hence, an equation relating the observer's radius $r_o$ to the star's radius $r_s$ is obtained. 

The equation~\eqref{alphas} with $r_m$ given by~\eqref{alpham} and applied to the observer with $r=r_o$, gives the angular size as dependent on the value of $r_o$.
By using both equations, the shadow angular size $\Tilde{\alpha}$, can be modeled with respect to the observer's radius $r_o$, as $r_s$ evolves. 

The beginning of the third stage occurs when $r_s=r_s^{(2)}$ or in terms of the minimum radius $r_m=r_{mp}$. The angular size will be given then by equation~\eqref{alpham} in the form:
\begin{equation}
    \sin{\Tilde{\alpha}}=\frac{\sqrt{1-2M(r_o)/r_o}\sqrt{\mathcal{R}(r_o,r_{mp})}}{\varepsilon\pm\sqrt{\varepsilon^2-1+2M(r_o)/r_o}\sqrt{1-\mathcal{R}(r_o,r_{mp})}}
\end{equation}
The sign plus is used until the observer crosses $r_o=r_{mp}$ after which the negative sign must be used. 
This change of sign is due to the change of sign in equation \eqref{rphirm} induced when $r=r_{mp}$.
This stage will last until the observer reaches $r_o=r_+$. 

\begin{figure}
\centering
\subfloat[$q_*=0.0$]
{\includegraphics[scale=0.45]{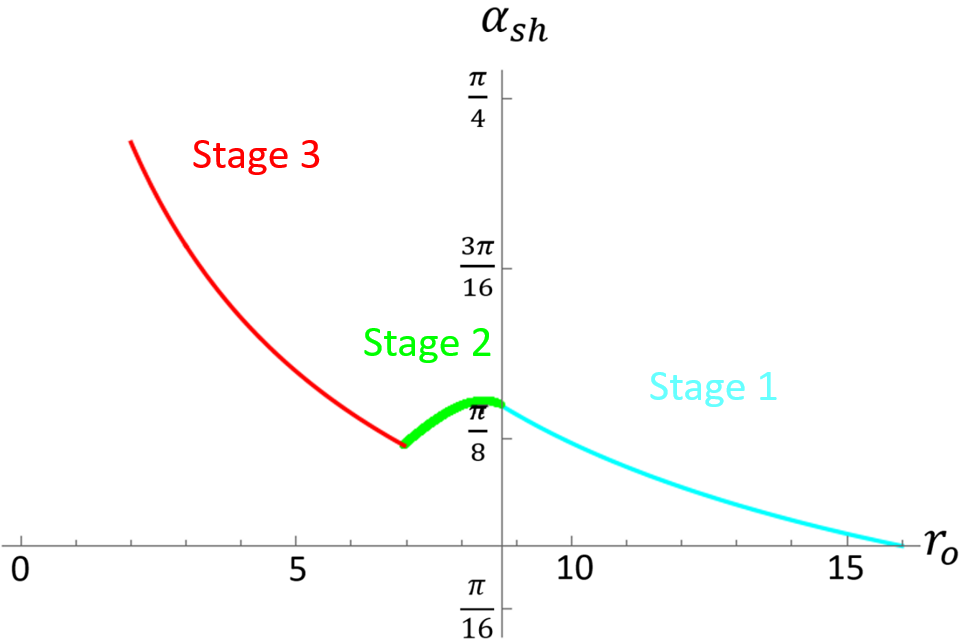}\includegraphics[scale=0.3]{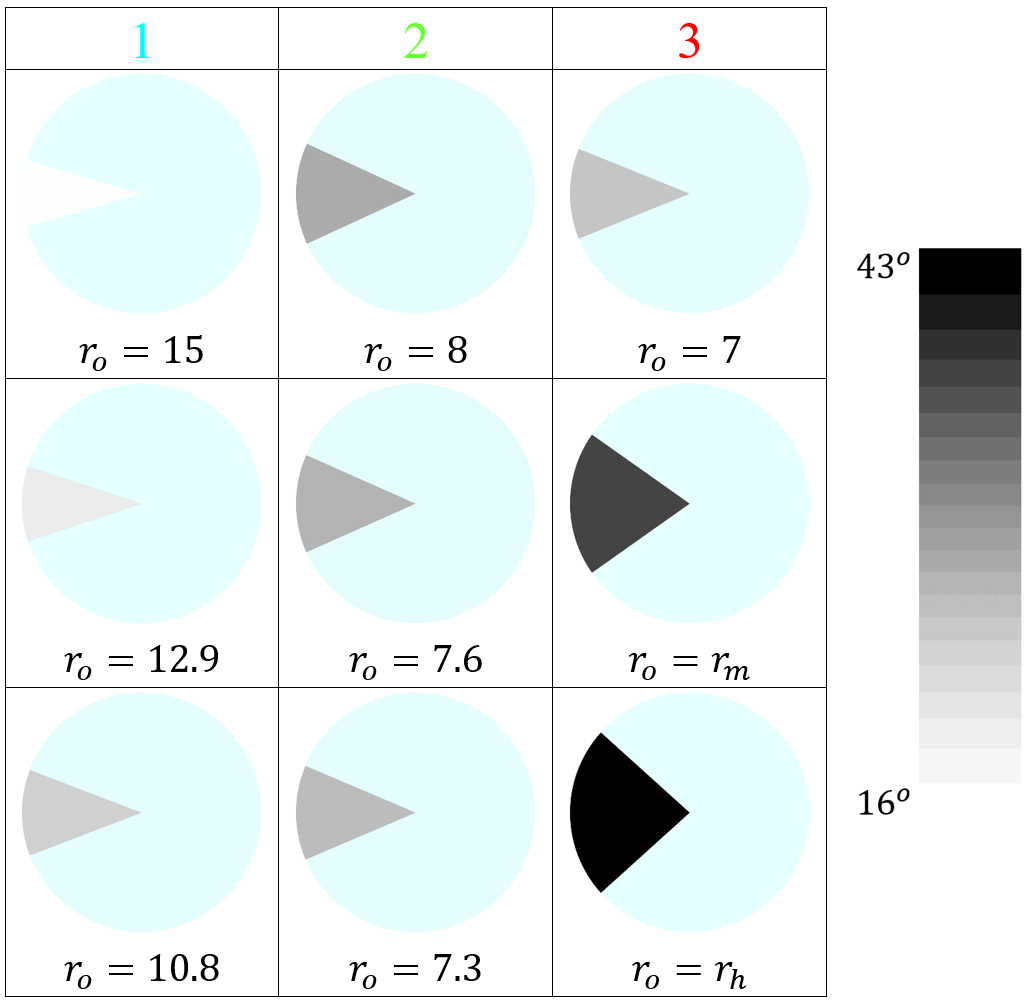}}\\
\subfloat[$q_*=0.8967$]
{\includegraphics[scale=0.45]{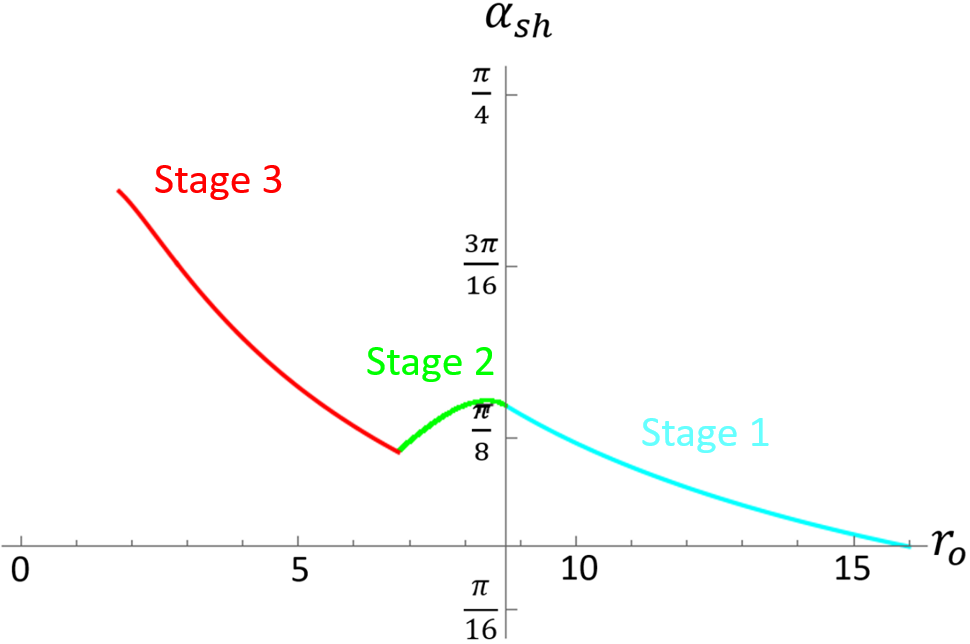}\includegraphics[scale=0.30]{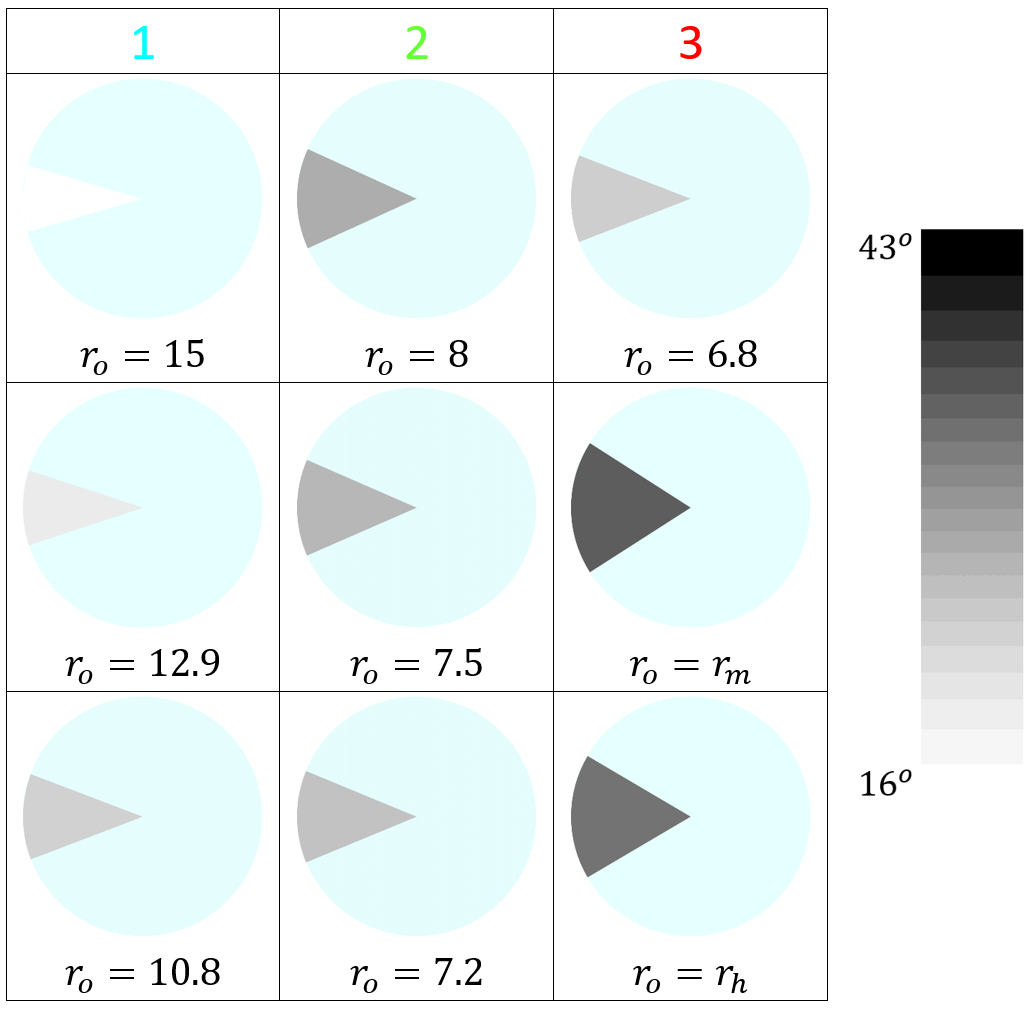}}\\
\subfloat[$q_*=1.0582$]
{\includegraphics[scale=0.45]{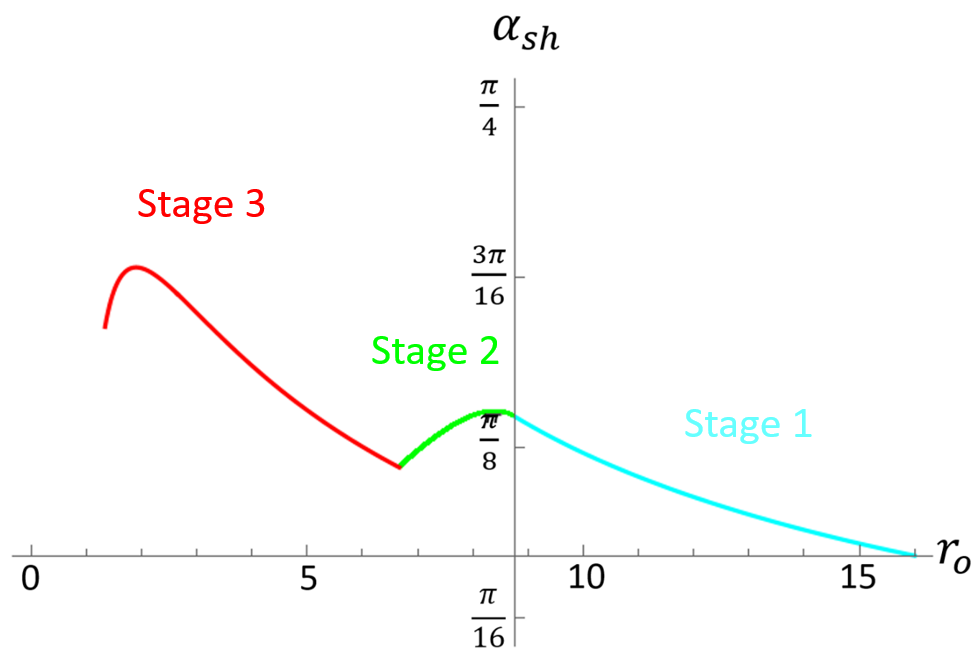}\includegraphics[scale=0.30]{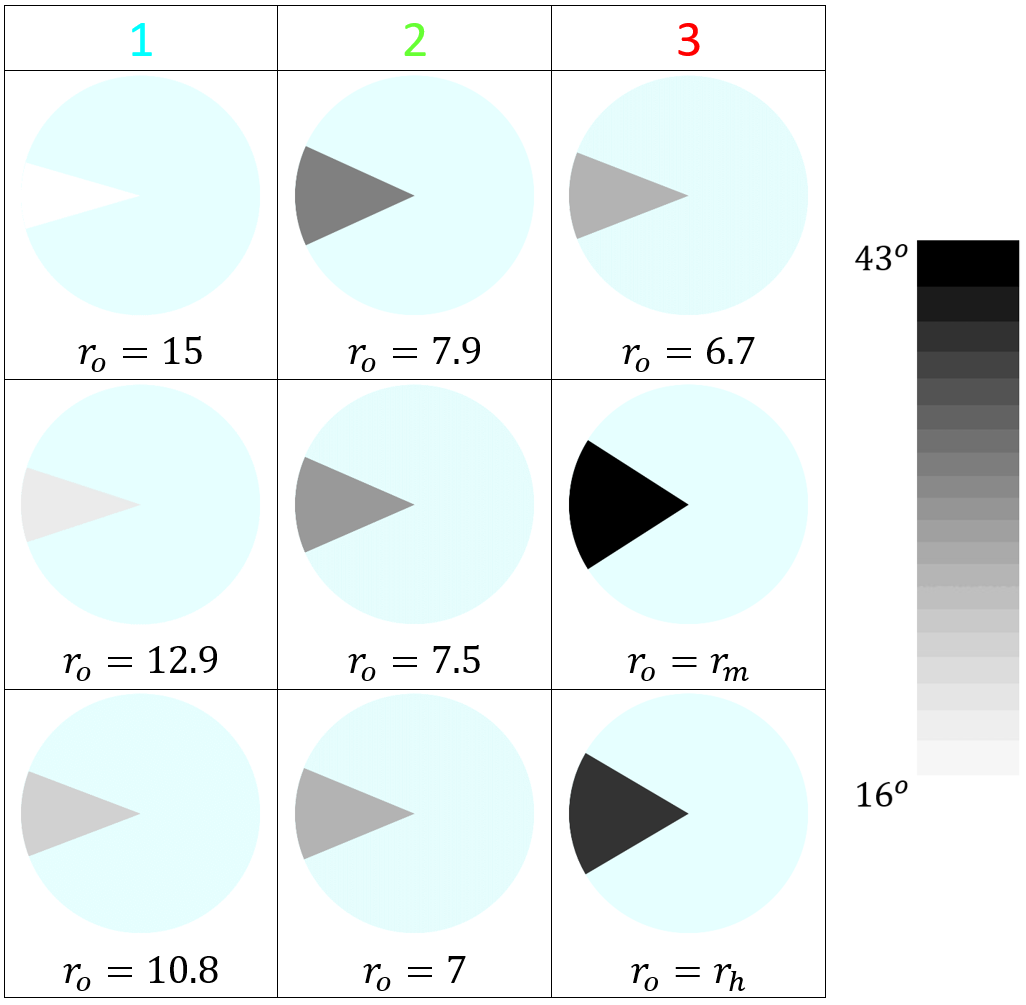}}
\caption{Plots of the spherical collapse of dust for different values of $q_*$. On the left side, the angular size $\alpha_{sh}$ is displayed as a function of the observer's radius $r_o$, on the right side, the representation of the viewer's sky are shown for different radii $r_o$. The corresponding  stage of collapse is indicated at the top of the figure.}
\label{gr3fases2}
\end{figure}

In the left panel of first row of figure~\ref{gr3fases2} the angular size of the shadow is plot as a function of the radius of the free falling observer for the collapse leading to a Schwarzschild black hole. 
In the plot, the three stages are shown in color. As the observer approaches the center, from the initial value $r_o^*=15$, the angular size increases monotonically up to a certain radius at which the size starts to decrease until a certain radius which characterizes the end of the second phase. After this radius the angular size increases again.
In the right panel of the first row,
the angular size is represented in the observer's sky. The three stages are represented
with the same color code as the left column.

The magnitude of the angular size is represented in the grey color scale that goes from $\tilde \alpha=16^\circ$ (white) to $\tilde\alpha=43^\circ$ (black).
From these figures, 
one can conclude that the effect on the shadow
of the Hayward parameter $q*$
becomes more noticeable as the collapse progresses. 
Furthermore, the beginning of the third stage occurs at smaller radius because the radius of the photosphere decreases with larger values of $q_*$. Therefore, the shadow have a smaller size at the end of the second stage. 
Additionally, at the end of the third stage, the shadow angle will be smaller, if the Hayward parameter is large enough, and more importantly it will have a local maximum. 
This behavior is related with the repulsive effect of the central core in the Hayward spacetime that damps the gravitational contraction and is manifested in the shadow measured by the infalling observer.
The values for the maximum in the third stage for a free falling observer are different depending in the value of the Hayward charge; for $q_*=0.0$ the maximum is at $\alpha_{sh}=0.734844$ for $r_s=2.0$, for $q_*=0.8967$ the maximum is at $\alpha_{sh}=0.675211$ for $r_s=1.76981$ and for $q_*=1.0582$ the maximum is at $\alpha_{sh}=0.600484$ for $r_s=1.90294$.
\section{Final remarks}
\label{sec:final}

In this work, we have analysed the formation of the shadow in a regular Hayward black hole. The evolution of the angular size of the shadow  during the collapse of a star in the regular Hayward space-time was shown by analyzing the null geodesics. 
Our results show that for a static observer outside the star, the behavior of the shadow essentially divides into three stages. The first and third stages maintain a constant shadow angle since the minimum radius $r_m$ remains constant, obeying equation~\eqref{rmrs}. The change in the shadow angular size occurs in the second stage where the star's radius $r_s$, as well as the minimum radius $r_m$, decreases in time due to the collapse of the star. The end of the second stage and the beginning of the third depend on the value of the Hayward parameter $q_*$, since the radius of the photo sphere depends on it. For a higher value of $q_*$, the second stage ends later, causing the shadow angle to reach a lower value. For observers between the photo sphere radius and the initial radius of the star, the evolution of the shadow begins in the second stage once the star's radius is smaller than that of the observer. For an observer between the horizon radius and the photo sphere radius, the value of the shadow angle is constant and depends on the radius of the photon sphere.
In the case of a radially falling observer, there are also three stages depending on the value of the minimum radius $r_m$. Unlike a static observer, this type of observer sees a change in the shadow angle at the end of the third stage. The relationship between the Hayward charge $q_*$ and the shadow angle is that as the Hayward charge increases, the shadow angle will decrease close to the end of the third stage. This decreasing in the angle is caused by the Hayward charge near $r=0$ that prevents the formation of the singularity during the collapse.
This study provides a perspective of the collapse evolution for regular black holes, revealing the effect of considering a regular metric to study the shadow of the collapsing star in such a regular Hayward spacetime.

\bibliographystyle{unsrt}
\bibliography{referencias} 

\begin{thebibliography}{10}

\bibitem{EventHorizonTelescope:2019ths}
Kazunori Akiyama et~al.
\newblock {First M87 Event Horizon Telescope Results. IV. Imaging the Central
  Supermassive Black Hole}.
\newblock {\em Astrophys. J. Lett.}, 875(1):L4, 2019.

\bibitem{EventHorizonTelescope:2019pgp}
Kazunori Akiyama et~al.
\newblock {First M87 Event Horizon Telescope Results. V. Physical Origin of the
  Asymmetric Ring}.
\newblock {\em Astrophys. J. Lett.}, 875(1):L5, 2019.

\bibitem{EventHorizonTelescope:2022wkp}
Kazunori Akiyama et~al.
\newblock {First Sagittarius A* Event Horizon Telescope Results. I. The Shadow
  of the Supermassive Black Hole in the Center of the Milky Way}.
\newblock {\em Astrophys. J. Lett.}, 930(2):L12, 2022.

\bibitem{EventHorizonTelescope:2022apq}
Kazunori Akiyama et~al.
\newblock {First Sagittarius A* Event Horizon Telescope Results. II. EHT and
  Multiwavelength Observations, Data Processing, and Calibration}.
\newblock {\em Astrophys. J. Lett.}, 930(2):L13, 2022.

\bibitem{EventHorizonTelescope:2022urf}
Kazunori Akiyama et~al.
\newblock {First Sagittarius A* Event Horizon Telescope Results. V. Testing
  Astrophysical Models of the Galactic Center Black Hole}.
\newblock {\em Astrophys. J. Lett.}, 930(2):L16, 2022.

\bibitem{Vagnozzi:2022moj}
Sunny Vagnozzi et~al.
\newblock {Horizon-scale tests of gravity theories and fundamental physics from
  the Event Horizon Telescope image of Sagittarius A}.
\newblock {\em Class. Quant. Grav.}, 40(16):165007, 2023.

\bibitem{Schneider:2018hge}
Stefanie Schneider and Volker Perlick.
\newblock {The shadow of a collapsing dark star}.
\newblock {\em Gen. Rel. Grav.}, 50(6):58, 2018.

\bibitem{Oppenheimer:1939ue}
J.~R. Oppenheimer and H.~Snyder.
\newblock {On Continued gravitational contraction}.
\newblock {\em Phys. Rev.}, 56:455--459, 1939.

\bibitem{Synge:1966okc}
J.~L. Synge.
\newblock {The Escape of Photons from Gravitationally Intense Stars}.
\newblock {\em Mon. Not. Roy. Astron. Soc.}, 131(3):463--466, 1966.

\bibitem{Hayward:2005gi}
Sean~A. Hayward.
\newblock {Formation and evaporation of regular black holes}.
\newblock {\em Phys. Rev. Lett.}, 96:031103, 2006.

\bibitem{Frolov:2016pav}
Valeri~P. Frolov.
\newblock {Notes on nonsingular models of black holes}.
\newblock {\em Phys. Rev. D}, 94(10):104056, 2016.

\bibitem{Flachi:2012nv}
Antonino Flachi and Jos\'e P.~S. Lemos.
\newblock {Quasinormal modes of regular black holes}.
\newblock {\em Phys. Rev. D}, 87(2):024034, 2013.

\bibitem{Lin:2013ofa}
Kai Lin, Jin Li, and Shuzheng Yang.
\newblock {Quasinormal Modes of Hayward Regular Black Hole}.
\newblock {\em Int. J. Theor. Phys.}, 52:3771--3778, 2013.

\bibitem{Lopez:2018aec}
L.~A. Lopez and Valeria Hinojosa.
\newblock {Quasinormal modes of Charged Regular Black Hole}.
\newblock {\em Can. J. Phys.}, 99(1):44--48, 2021.

\bibitem{Chiba:2017nml}
Takeshi Chiba and Masashi Kimura.
\newblock {A note on geodesics in the Hayward metric}.
\newblock {\em PTEP}, 2017(4):043E01, 2017.

\bibitem{Wei:2015qca}
Shao~Wen Wei, Yu~Xiao Liu, and Chun~E. Fu.
\newblock {Null Geodesics and Gravitational Lensing in a Nonsingular
  Spacetime}.
\newblock {\em Adv. High Energy Phys.}, 2015:454217, 2015.

\bibitem{Zhao:2017cwk}
Shan-Shan Zhao and Yi~Xie.
\newblock {Strong deflection gravitational lensing by a modified Hayward black
  hole}.
\newblock {\em Eur. Phys. J. C}, 77(5):272, 2017.

\bibitem{Kumar:2019pjp}
Rahul Kumar, Sushant~G. Ghosh, and Anzhong Wang.
\newblock {Shadow cast and deflection of light by charged rotating regular
  black holes}.
\newblock {\em Phys. Rev. D}, 100(12):124024, 2019.

\bibitem{AbhishekChowdhuri:2023ekr}
Abhishek Chowdhuri, Saptaswa Ghosh, and Arpan Bhattacharyya.
\newblock {A review on analytical studies in Gravitational Lensing}.
\newblock {\em Front. Phys.}, 11:1113909, 2023.

\bibitem{Abbas:2014oua}
G.~Abbas and U.~Sabiullah.
\newblock {Geodesic Study of Regular Hayward Black Hole}.
\newblock {\em Astrophys. Space Sci.}, 352:769--774, 2014.

\bibitem{Bautista-Olvera:2019blb}
Brandon Bautista-Olvera, Juan~Carlos Degollado, and Gabriel German.
\newblock {Geodesic structure of a rotating regular black hole}.
\newblock {\em Gen. Rel. Grav.}, 55(5):66, 2023.

\bibitem{Debnath:2015hea}
Ujjal Debnath.
\newblock {Accretion and Evaporation of Modified Hayward Black Hole}.
\newblock {\em Eur. Phys. J. C}, 75:129, 2015.

\bibitem{Carballo-Rubio:2018pmi}
Ra\'ul Carballo-Rubio, Francesco Di~Filippo, Stefano Liberati, Costantino
  Pacilio, and Matt Visser.
\newblock {On the viability of regular black holes}.
\newblock {\em JHEP}, 07:023, 2018.

\bibitem{Carballo-Rubio:2018jzw}
Ra\'ul Carballo-Rubio, Francesco Di~Filippo, Stefano Liberati, and Matt Visser.
\newblock {Phenomenological aspects of black holes beyond general relativity}.
\newblock {\em Phys. Rev. D}, 98(12):124009, 2018.

\bibitem{Perez-Roman:2018hfy}
Ivan Perez-Roman and Nora Bret\'on.
\newblock {The region interior to the event horizon of the Regular Hayward
  Black Hole}.
\newblock {\em Gen. Rel. Grav.}, 50(6):64, 2018.

\end{thebibliography}

\end{document}